%% file: main.tex
\documentclass[compsoc,conference,a4paper,10pt,times]{IEEEtran}
\IEEEoverridecommandlockouts
\usepackage{cite}
\usepackage{amsmath,amssymb,amsfonts}
\usepackage{algorithm}
\usepackage[noend]{algpseudocode}

\usepackage{graphicx}
\usepackage{textcomp}
\usepackage{bmpsize}
\usepackage{xcolor}
\usepackage{colortbl}
\usepackage{lipsum}
\usepackage[colorlinks=true,urlcolor=black]{hyperref}
\def\BibTeX{{\rm B\kern-.05em{\sc i\kern-.025em b}\kern-.08em
    T\kern-.1667em\lower.7ex\hbox{E}\kern-.125emX}}

\usepackage{array}
\usepackage{gensymb}
\usepackage{booktabs}
\usepackage{xspace}
\usepackage{bbm}
\usepackage[caption=false,font=footnotesize]{subfig}
\usepackage{comment}
    
\newtheorem{thm}{Theorem}

\DeclareMathOperator{\E}{\mathbb{E}}

\DeclareMathOperator*{\argmax}{argmax}
\DeclareMathOperator*{\argmin}{argmin}

\newcommand{\alg}{GRAPHITE\xspace}

\newcommand{\target}{y_{tar}}
\newcommand{\etal}{\emph{et al}.\xspace}

\newcommand{\eotmetric}{transform-robustness\xspace}
\newcommand{\EOTmetric}{Transform-robustness\xspace}
\newcommand{\MYhref}[3][red]{\href{#2}{\color{#1}{#3}}}

\def\includedriveby{0}
\def\arxiv{0}

\begin{document}

%\title{\alg: A Practical Framework for Generating Automatic Physical Adversarial Machine Learning Attacks}
\title{\alg: Generating Automatic Physical Examples for Machine-Learning Attacks on Computer Vision Systems}

%\author{Ryan Feng${}^1$, Neal Mangaokar${}^1$, Jiefeng Chen${}^2$, Earlence Fernandes${}^2$, Somesh Jha${}^2$, Atul Prakash${}^1$\\
%{\normalsize ${}^1$University of Michigan, Ann Arbor, ${}^2$University of Wisconsin, Madison}\\
%{\normalsize ${}^1$\{rtfeng, nealmgkr, aprakash\}@umich.edu, ${}^2$\{jiefeng, earlence, jha\}@cs.wisc.edu}}

 \author{\IEEEauthorblockN{Ryan Feng${}^1$, Neal Mangaokar${}^1$, Jiefeng Chen${}^2$, Earlence Fernandes${}^2$, Somesh Jha${}^2$, Atul Prakash${}^1$}
 \IEEEauthorblockA{\normalfont ${}^1$University of Michigan, Ann Arbor, ${}^2$University of Wisconsin, Madison \\
 \normalfont ${}^1$\{rtfeng, nealmgkr, aprakash\}@umich.edu, ${}^2$\{jiefeng, earlence, jha\}@cs.wisc.edu
 }}
% \and
% \and
% \and
% \and
% \and

\maketitle

\begin{abstract}
This paper investigates an adversary's ease of attack in generating adversarial examples for real-world scenarios. We address three key requirements for practical attacks for the real-world: 1) \emph{automatically} constraining the size and shape of the attack so it can be applied with stickers, 2) {\em transform-robustness}, i.e., robustness of a attack to environmental \emph{physical} variations such as viewpoint and lighting changes, and 3) supporting attacks in not only white-box, but also black-box \emph{hard-label} scenarios, so that the adversary can attack proprietary models. In this work, we propose \alg, an efficient and general framework for generating attacks that satisfy the above three key requirements.  \alg takes advantage of transform-robustness, a metric based on expectation over transforms (EoT), to \emph{automatically} generate small masks and optimize with gradient-free optimization. \alg is also flexible as it can easily trade-off transform-robustness, perturbation size, and query count in black-box settings. On a GTSRB model in a hard-label black-box setting,  we are able to find attacks on all possible 1,806 victim-target class pairs with averages of 77.8\% transform-robustness, perturbation size of 16.63\% of the victim images, and 126K queries per pair. For digital-only attacks where achieving transform-robustness is not a requirement, \alg is able to find successful small-patch attacks with an average of only 566 queries for 92.2\% of victim-target pairs. \alg is also able to find successful attacks using perturbations that modify small areas of the input image against PatchGuard, a recently proposed defense against patch-based attacks.

\end{abstract}


\input{tex/intro.tex}

\input{tex/related_work.tex}
\input{tex/setup}

\input{tex/whitebox.tex}

\input{tex/blackbox.tex}

\input{tex/experiments.tex}

\input{tex/conclusion.tex}

\section*{Acknowledgements}
This material is based on work supported by DARPA under agreement number 885000, Air Force Grant FA9550-18-1-0166, the National Science Foundation (NSF) Grants 1646392, 2039445, CCF-FMitF-1836978, SaTC-Frontiers-1804648 and CCF-1652140, and ARO grant number W911NF-17-1-0405. Earlence Fernandes is supported by the University of Wisconsin-Madison Office of the Vice Chancellor for Research and Graduate Education with funding from the Wisconsin Alumni Research Foundation. The U.S. Government is authorized to reproduce and distribute reprints for Governmental purposes notwithstanding any copyright notation thereon. Any opinions, findings, and conclusions or recommendations expressed in this material are those of the author(s) and do not necessarily reflect the views of our research sponsors. We thank Nelson Manohar for his help in early development, and Renuka Kumar, Washington Garcia, David Fouhey, Pin-Yu Chen, and Varun Chandrasekaran for their feedback.

{
\bibliographystyle{IEEEtran}
\bibliography{references}
}

\appendices

\section{NP-Completeness of Mask Generation}\label{ap:mask_gen}
\input{tex/np_complete}

\section{Transformation Details}\label{ap:xforms}
\input{tex/transform_details}

\section{Additional White-box Details and Results}\label{ap:wb_hyper}
\input{tex/wb_hyper}

\section{Additional CIFAR-10 Results}\label{ap:cifar}
\input{tex/cifar}

\section{Drive-by Test Images}\label{ap:drive-by}
\input{tex/drive_by}

\section{Printing and Lighting Error}\label{ap:blue}
\input{tex/blue}

\section{ALPR Attack}\label{ap:alpr}
\input{tex/alpr}

\end{document}

%% file: tex/intro.tex
\section{Introduction}
Machine learning (ML) models have had resounding success in several scenarios such as face and object recognition~\cite{he2016deep,krizhevsky2012imagenet,schroff2015facenet,simonyan2014very,szegedy2016rethinking}. Therefore, such models are now a part of perception pipelines in cyber-physical systems like cars~\cite{geiger2012we,lillicrap2015continuous,openpilot}, UAVs~\cite{bou2010controller,mostegel2016uav} and robots~\cite{zhang2015towards}. However, recent work has found that these models are vulnerable to subtly perturbed adversarial examples that cause misclassification~\cite{carlini2017towards,goodfellow2014explaining,papernot2016limitations,szegedy2013intriguing}. While these early digital white-box attacks were useful in understanding model weaknesses, researchers are now considering how to make attacks more practical for adversaries to accomplish in the real world~\cite{athalye2017synthesizing,patch,roadsigns17,glasses,cheng2018query,brendel2017decision}.

Practical, real-world attacks have three key requirements. The first is that they should \emph{automatically} choose small areas to perturb, so that they can be applied with stickers on existing physical objects. The second is that practical, real-world attacks should exhibit {\em \eotmetric}, i.e., be robust to \emph{physical}-world effects such as viewpoint and lighting changes as measured by a metric based on expectation over transforms~\cite{athalye2017synthesizing}. The final requirement is that the most powerful real-world attacks should only require \emph{hard-label} access. This means that the adversary can succeed with only the final top-1 decision, which \emph{must} be provided by a deployed ML model for it to be useful even if its internals are closed source and designed to be difficult to completely reconstruct due to proprietary formats, implementations, and stripped binaries~\cite{keen-lab}. The adversary should also be able to work with smaller query budgets when that is a constraint.  

Regarding the first requirement, the art of \emph{automatically} picking limited areas to perturb remains relatively unexplored. Existing patch attacks~\cite{patch,lee2019physical,liu2018dpatch,fawzi2016measuring,yang2020patchattack} limit perturbations to small patches, but are restricted to pre-defined patch shapes. The adversary must also either search for a good location and size or optimize over an expectation of patch locations and sizes. This is inefficient and, in black-box settings, query-intensive. Existing work also shows that targeted attacks require larger patches~\cite{yang2020patchattack}. RP${}_2$~\cite{roadsigns17} uses masks to constrain the space of the attack, but the mask creation process requires some manual experimentation, making it harder to scale. The Carlini \& Wagner $\ell_0$ attack~\cite{carlini2017towards} generates arbitrary, small sets of pixels to perturb, but is designed for digital attacks, not for physical settings or where only hard-label access to a model is available (requirements 2 and 3). 

Likewise, existing work on finding attacks in the hard-label case do not satisfy the first two requirements; instead, they find digital attacks that require tens or hundreds of thousands of queries~\cite{cheng2018query, brendel2017decision, cheng2019signopt, chen2019hopskipjumpattack}.  %For the first time (to our knowledge), we combine all three key requirements to create a more practical real-world attack framework.

We generalize these principles and \emph{combine all three requirements} for the first time, to our knowledge, creating a more practical and efficient real-world attack framework called \alg (\underline{G}enerating \underline{R}obust \underline{A}utomatic \underline{PH}ysical \underline{I}mage \underline{T}est \underline{E}xamples). Such an attack framework could be potentially useful for security testing of and improving future defenses. We formulate our automatic framework of physical attacks as a joint optimization problem, balancing the opposing goals of minimizing perturbation size (or ``mask size'') and maximizing \eotmetric. More broadly, the framework is designed to find attacks that permit tradeoffs among different constraints of query budgets (in hard-label settings), mask size, and \eotmetric. 

We  first instantiate this framework in the white-box setting by adapting the Carlini-Wagner $\ell_0$ attack~\cite{carlini2017towards} to help satisfy the three requirements. We show that the resulting attacks require a nearly 2x lower number of forward/backward passes compared to typical square patch attacks and exhibit more \eotmetric.  

We then propose an instantiation of the framework for the more challenging hard-label setting.  We show that there are significant algorithmic differences between the white-box and black-box attacks that are motivated by the general difficulties of na\"{i}ve $\ell_0$ style minimization attacks in the highly discrete and discontinuous hard-label optimization space. As a result, we show experimentally that simply extending white-box attacks to hard-label black-box settings using strategies such as OPT~\cite{cheng2018query} yields poor results in Section~\ref{sec:baselines}.
%For instance, one of the challenges is that hard-label gradients fail to create a reliable importance ordering of pixels in EoT settings. The second challenge is that hard-label reformulation algorithms such as OPT-attack~\cite{cheng2018query} require smooth functions with the ability to sample slightly different values in a nearby neighborhood. Such techniques also require finding a initial point at which the objective changes in nearby neighborhoods.  
We solve these challenges by separating the joint optimization into a two-stage approximation: automatic mask generation and perturbation boosting, while using transform-robustness to provide a more continuous measure (as compared to just success or failure of an attack) to determine the relative value of keeping or removing pixels. 

%In mask generation, we start with the target image, estimate a heatmap with \eotmetric measures on prospective patch removals, and reduce the image according to a mask scoring function, giving us a principled way to reduce the set of pixels in the mask while ensuring a reasonably transform-robust initialization. Then, we show empirically that directly optimizing \eotmetric is sufficiently smooth for gradient-free optimization and boost the \eotmetric of the perturbation as high as we can for the chosen mask. 
%The switch from a jointhard label gradients don't create a reliable heatmap? Digital $\ell_2$ or $\ell_\infty$ perturbation attacks solve this by exploring directions near and around the decision boundary~\cite{brendel2017decision,chen2019hopskipjumpattack,cheng2018query,cheng2019signopt}. However, in the $\ell_0$ case, which is non-differentiable, there is no clear definition of a direction to reduce the $\ell_0$ perturbation - such a problem would be combinatorially hard. Furthermore, with attacks such as OPT-attack~\cite{cheng2018query}, the problem is reformulated as one that minimizes distance so that the objective becomes somewhat smooth. This objective becomes infeasible to solve when transformations are added in, as we show in Section~\ref{sec:exp}.

%To solve these problems, we utilize ``\emph{\eotmetric}" to automatically identity useful masks and then find optimal perturbations to place in those masks. 

 \begin{figure}[t]
     \centering
       \subfloat[Targeted GTSRB attack to convert a \texttt{Stop sign} into a \texttt{Speed Limit 30 km/hr sign}.\label{fig:intro_gtsrb}]{\includegraphics[width=1.182in]{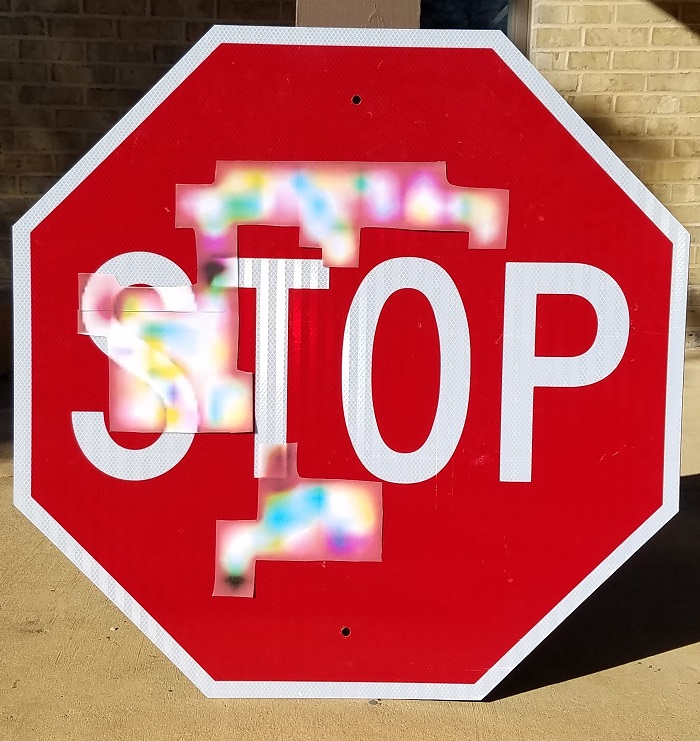}}%
       \hfil
       \subfloat[Targeted CIFAR-10 attack that  modifies just 10 pixels to defeat the recently proposed PatchGuard~\cite{chong2021patchguard} defense to misclassify a dog to a cat. \label{fig:intro_cifar}]{\includegraphics[width=1.25in]{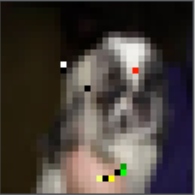}}%
    \caption{Example GTSRB and CIFAR-10 \alg-generated black-box hard-label attacks.}\label{fig:examples}
    \vspace{-0.1in}
\end{figure}

%We thus propose \alg (Generating Robust Adversarial PHysical Image Test Examples), a hard-label instantiation of our framework and the first pipeline to generate physically robust adversarial examples in the hard-label setting. We solve the above challenges by approximating the joint optimization framework with a two stage optimization process: mask generation and perturbation boosting. We propose a mask generation phase that performs our framework's pixel selection and removal by estimating \eotmetric heatmap and greedily reducing the image according to a mask scoring function. This gives us a principled way to \emph{automatically} reduce the set of pixels in the mask while ensuring a reasonably transformation robust initialization. We then show empirically that directly optimizing the \eotmetric of the remaining pixels is sufficiently smooth for gradient-free optimization~\cite{nesterov2017random,ghadimi2013stochastic} to solve, so we use this process to boost the \eotmetric of the perturbation as high as we can for the chosen mask.

We evaluate our \alg framework in hard-label black-box settings both digitally on all 1,806 victim-target pairs of German Traffic Sign Recognition Benchmark (GTSRB) classes, and physically with printed stickers placed on a real \texttt{Stop sign}. We add an additional constraint in these experiments to restrict the attack to be contained within the victim object for physical-realizability. Digitally, we achieve an average \eotmetric of 77.8\% and an average mask size of 16.63\% of the victim image with 126k queries. Physically, we ran tests over 420 field test images of printed stickers placed on a real \texttt{Stop sign}. This includes 210 images of a \texttt{Speed Limit 30 km/hr sign} attack with a 95.7\% attack success rate. We then demonstrate our framework's ability to trade off mask size, \eotmetric, and query count. In one such setting, when \eotmetric is not required or low values suffice, \alg is able to find successful small-patch attacks for over 92.2\% of GTSRB victim-target pairs with just 566 queries per successful pair. In contrast, state-of-the-art hard-label attacks of any type (including those bounded by $\ell_2$ or $\ell_\infty$ norms) on similar datasets all report requiring tens to hundreds of thousands of queries, i.e., one to three orders of magnitude higher.

Finally, to demonstrate \alg's ability to \emph{thwart existing defenses}, we select a state-of-the-art patch attack defense called PatchGuard~\cite{chong2021patchguard} and attack a defended CIFAR-10 model in the hard-label black-box setting. Across 100 random victim-target image pairs, we are able to successfully attack PatchGuard by perturbing as little as 1\% of the pixels (e.g., see Figure~\ref{fig:intro_cifar}). This also demonstrates the value of a framework that can generate more generalized forms of patch attacks (e.g., attacks using multiple patches) than those considered by PatchGuard; PatchGuard primarily evaluated its scheme on patches in a single part of an image.

The code is available at: \MYhref[magenta]{https://github.com/ryan-feng/GRAPHITE}{https://github.com/ryan-feng/GRAPHITE}.

\vspace{0.1in}
\noindent\textbf{Our Contributions:}

\begin{itemize}
    \item We proposed a general framework for generating practical, real-world attacks that combine three key requirements (that they are \emph{automatically} space constrained, \emph{physically} robust to transformations, and only require black-box \emph{hard-label} access). The algorithms based on the framework generate attacks efficiently in both white-box and black-box settings.   %In whitebox settings, The framework helps generate attacks with fewer forward and backward passes and higher \eotmetric through the model than the current state-of-the-art method of selecting a square patch and then optimizing it.   %We propose a white-box adaptation of existing $\ell_0$ attack methods to satisfy the first two requirements, and then explore the more challenging hard-label black-box scenario to also satisfy the third requirement.
    
    \item We introduce the first automatic, robust physical perturbation attacks in the black-box hard-label setting. We presented a novel mask generation process that \emph{automatically} generates candidate masks to constrain the perturbation size and showed that \eotmetric is a sufficiently smooth function to optimize with gradient-free optimization. We test both digitally and physically, with printed stickers placed on physical objects.
    
    \item We demonstrated \alg's ability to trade off \eotmetric, query budget, and mask size depending on the adversary's limitations and priorities. As an example, we generate attacks with non-negligible \eotmetric in 92.2\% of GTSRB victim-target pairs with an average of just 566 queries and an average mask size of 10.5\%.
    %algorithm for creating robust, physical adversarial examples to the best of our knowledge. We provide a formulation of a joint optimization problem to find optimal masks and perturbations and present algorithms to solve them using \eotmetric. 
    
    %\item % To the best of our knowledge, our attack is the first physical attack to generate masks automatically in either white-box or black-box settings.
    
    %\item %We evaluate effectiveness of GRAPHITE on multiple scenarios including attacks on models for classifying traffic signs as well as recognizing license plates (in the latter, just by perturbing the small area normally occupied by a license plate holder). We perform both digital and real-world field tests.
    
    %\item  We provide a formulation of a joint optimization problem across masks and robustness of the classification to environmental transforms. We show that the problem is computationally hard. We present algorithms to address it and include theoretical analysis of the algorithms.
    
    %\item We evaluate \alg on multiple (victim, target) pairs on the GTSRB dataset and found that it succeeded in finding perturbations with both small mask area and high robustness, based on expectation over transforms (EOT) metrics proposed by Athalye et al.~\cite{athalye2017synthesizing}. We also confirm the robustness of attacks in physical settings by applying stickers to a stop sign, as suggested by \alg, using the evaluation methodology of Eykholt et al.~\cite{roadsigns17}.
    
    % demonstrate that existing techniques are not satisfactory

\end{itemize}

%% file: tex/related_work.tex
\section{Related Work}
\textbf{Automatic Attacks.} The art of constraining the size and shape of perturbations remains relatively unexplored. One approach to limit the perturbation size is to use a patch attack. Patch attacks were first proposed as a white-box technique that crafted small perturbation patches that could be added to any scene to cause a misclassification~\cite{patch}. Later work then explored patch attacks on object detection~\cite{lee2019physical,liu2018dpatch} and the soft-label setting~\cite{fawzi2016measuring,yang2020patchattack}, where the model output probabilities are available. Oftentimes, the patches are made robust to different spatial locations by optimizing over different patch locations. However, these approaches must run inferences over many locations and are stuck to static shapes. In contrast, we are able to efficiently generate arbitrarily shaped perturbations and do so with only hard-label access, where only the top-1 class prediction is known.

Other work that limits the size of perturbation includes RP${}_2$~\cite{roadsigns17}, a physical white-box attack that designs adversarial stickers and the Carlini-Wagner (C\&W) $\ell_0$ attack~\cite{carlini2017towards}. RP${}_2$ introduces the notion of a \emph{mask} to limit the size of the perturbation. However, their process requires some manual experimentation based on the visual results of an $\ell_1$ approximation. The C\&W $\ell_0$ attack alternates between pixel removal and the C\&W $\ell_2$ attack to optimize an $\ell_0$, but is designed for digital white-box setting, failing to meet requirements 2 and 3.

\textbf{Physical Attacks.} Existing work for robust physical
perturbation attacks on image models remains in the \emph{white-box} setting. Examples include printing images of perturbed objects~\cite{kurakin2016adversarial}, modifying objects with stickers~\cite{roadsigns17,glasses}, and 3D printing perturbed objects~\cite{athalye2017synthesizing}. Such attacks typically optimize the expectation over a distribution of transformations, known as \textit{Expectation over Transformation (EoT)}~\cite{athalye2017synthesizing}. Such transformations model changes in viewing angle, distance, lighting, etc. In this work, we describe the expectation of attack success over such transformations as the \emph{\eotmetric} of an attack, approximated by measuring the attack success rate with some number of transformations. Some attacks also introduce the notion of a mask that constrains the perturbation to small regions of the object~\cite{roadsigns17}. However, existing work does not fully automatically generate suitable masks, and the mask is not jointly optimized with the noise.

In contrast, we automatically generate small masks to confine perturbations. We also extend our construction of such attacks to the black-box hard-label setting.

%To the best of our knowledge, our work is the first to demonstrate robust and query-efficient physical attacks using only top-1 hard-label information. 

%Furthermore, \alg, as far as we are aware, is the first to provide an automated method to both generate a small mask and find a robust adversarial example over transforms. Eykholt et al.~\cite{roadsigns17} came closest to that in whitebox settings, suggesting possible ways to identify sensitive areas in an image, but in the end, did not provide an automated method to generate a good mask.

 %Our work take inspiration from transformation  techniques in~\cite{roadsigns17}. We express physical transformations using classical computer vision techniques. For example, we model all geometric transformations (e.g., translation, rotation) using a homography matrix and model lighting changes with a radiometric transformation (see Appendix~\ref{subsec:transforms}).

\textbf{Black-box Attacks.} 
Recent work has explored \emph{digital} hard-label attacks~\cite{brendel2017decision,chen2019hopskipjumpattack,cheng2018query,cheng2019signopt,ilyas2018}, where only the top-1 label prediction is available. Such attacks are difficult because the hard-label optimization space is discontinuous and discrete. One approach is to reformulate the objective into a continuous optimization problem that finds the perturbation direction at which the decision boundary is closest. This approach is taken by OPT-attack~\cite{cheng2018query}, which uses the Randomized Gradient Free (RGF) method~\cite{ghadimi2013stochastic,nesterov2017random} to optimize this distance-based objective. The boundary distance is calculated through a binary search process. The perturbation is initialized by interpolating with a target class image. 

In our work, we generate physically robust examples within small, masked areas with only hard-label access. As we show in Section~\ref{sec:baselines}, directly adding EoT to this distance-based reformulation is both inefficient and ineffective, leaving visible artifacts of the intended target. Unlike OPT-attack, we also generate masks. 

Other black-box work includes transfer and soft-label attacks. Transfer attacks train a surrogate model, generate white-box examples on the surrogate, and hope they transfer to the target~\cite{papernot2016transferability}. Unfortunately, targeted examples often fail to transfer~\cite{delving}. Many techniques also require access to similar training sets that may not be available, whereas our work only requires query access to the target model. Soft-label attacks, or \textit{score-based} attacks, require access to the softmax layer output in addition to class labels~\cite{chen2017zoo,narod-transfer}. In contrast, our threat model only allows the adversary top-1 predicted class label access.

%\noindent\textbf{Black-box physical attacks on audio.}
%There is recent work in the audio domain, and specifically automatic speech recognition (ASR) systems, that uses the threat model of black-box physical attacks~\cite{hiddenvoicecommands,noodles,devil}. One such approach is based on mangling voice commands based on reverse engineering MFCC features~\cite{hiddenvoicecommands,noodles}, which is specific to the ASR domain. Other approaches build upon the transferability principle~\cite{devil}, where the attacker trains a surrogate model, performs white-box attacks on that model, and then hopes that the attacks transfer to the target. As discussed earlier, our approach does not require a dataset or training a surrogate model; furthermore, adversarial examples based on transferability do not always transfer to the target model, especially when conducting targeted attacks~\cite{chen2017zoo,narod-transfer}. In contrast, our work is robust on targeted examples and requires only the model's top-1 predicted label without needing to train a surrogate model. Additionally, robust physical audio adversarial examples with low sound distortion appear to still be an open challenge in the white-box setting~\cite{qin-audio}. By contrast, our work builds a structured approach in the more established area of vision adversarial examples where white-box attacks are already physically robust~\cite{athalye2017synthesizing,roadsigns17,yolo} and adapt to the black-box setting.

%% file: tex/setup.tex
\section{Setting up Automatic Physical Attacks}
In this section, we lay the groundwork for our general joint optimization problem for automatic physical attacks. We define automatic physical attacks to be ones that: 1) automatically pick a small mask consisting of pixels to perturb and 2) are robust to some set of transformations, set to model environmental variation such as viewpoint and lighting changes. We describe this joint optimization problem as the ideal problem we strive for and will provide solvers for in Sections~\ref{sec:white-box} and~\ref{sec:black-box}. We also show that mask generation is an NP-complete problem. 

%-------------------------------------------------------------------------
\subsection{Problem Setup}
Our goal is to find some perturbation $\delta$ and a small mask $M$ such that when $\delta$ is applied to a victim image $x$ in the area defined by mask $M$, our model $F$ predicts the target label $\target$ with high \eotmetric.  \EOTmetric is estimated over $t$ transforms sampled from a distribution of transforms $T$.  We describe a targeted attack formulation but can easily adapt to untargeted.

%We also introduce a general mask goodness function $h(M)$ that rewards masks that have configurable desirable properties, such as masks with a guaranteed minimum \eotmetric or masks that have a certain shape.

%-------------------------------------------------------------------------

% \usepackage{booktabs}
\begin{table*}
\centering
\caption{A summary of characteristics between all baselines and \alg. We adapt C\&W $\ell_0$ to white-box physical attacks and to black-box physical attack baselines. Upon finding these baselines to be insufficient, we split the joint optimization into a two-step optimization problem with a single round of mask generation and boosting.}\label{tab:comparisons}
\begin{tabular}{@{}cccccc@{}}
\toprule
\bf Algorithm & \bf \begin{tabular}[c]{@{}c@{}}Works in \\ Hard-Label?\end{tabular} & \bf Uses EoT?  & \bf \begin{tabular}[c]{@{}c@{}}Pixel Removal \\ Strategy\end{tabular}  & \bf \begin{tabular}[c]{@{}c@{}}Internal Attack\end{tabular} & \bf \begin{tabular}[c]{@{}c@{}}Num. Pixel Removal\\ + Attack Rounds\end{tabular}   \\ \midrule
C\&W $\ell_0$~\cite{carlini2017towards}   &             &                                &      Gradient-based                                                                                                                                       &     C\&W $\ell_2$~\cite{carlini2017towards}            &        Multiple                                \\
\begin{tabular}[c]{@{}c@{}}White-box \alg (Sec.~\ref{sec:white-box})\end{tabular}        &             &  \checkmark                              &       Gradient-based                                                                                                                                         &         EoT PGD        &                  Multiple                        \\
 $\ell_0$ and OPT (Sec.~\ref{subsec:l0opt})        &     \checkmark        &                                   &        RGF-based                                                                                                                                       & OPT~\cite{cheng2018query}                &                                 Multiple        \\
 $\ell_0$ and OPT w/ EoT (Sec.~\ref{subsec:l0opteot})       &    \checkmark         &          \checkmark                       &           RGF-based                                                                                                                                    &    OPT w/ EoT             &          Multiple                                \\
 $\ell_0$ and Boosting (Sec.~\ref{subsec:l0boost})        &      \checkmark       &             \checkmark                    &          RGF-based                                                                                                                                    &  Boosting (Sec.~\ref{sec:boost})              &               Multiple                \\
 \begin{tabular}[c]{@{}c@{}}Black-box \alg (Sec.~\ref{sec:bb_graphite})\end{tabular}         &     \checkmark        &               \checkmark                  &           Mask Gen. (Sec.~\ref{sec:mask_gen})                                                                                                                                 &      Boosting (Sec.~\ref{sec:boost})          &           1                             \\ \bottomrule
 \vspace{-0.2in}
\end{tabular}
\end{table*}
%\vspace{-0.1in}

\subsection{Optimizing Mask and \EOTmetric}\label{sec:joint_optimization_formulation}
The optimal perturbation maximizes \eotmetric (i.e., EoT metric) while occupying only a small area of the object such as a traffic sign.  $M_{ij}$ is 1 if pixel at position $(i,j)$ is perturbed and 0 otherwise. A way to model such constrained optimization problems is to model is as a joint optimization problem, which we give below. $\lambda  > 0$ is the relative weight given to size of the mask $M$ in the joint objective:

%\vspace{-0.1in}
\begin{equation}\label{eq:joint}
    \begin{aligned}
        \argmin_{\delta, M} \; & \lambda \cdot \vert\vert M\vert\vert_0 %+ \lambda_2 \cdot h(M) 
        - \E_{t \sim T} \bigg[ F\bigg(t(x + M \cdot \delta)\bigg) = \target \bigg]\\
        %\textrm{s.t.} \quad & \forall_i \hspace{0.2em} M[i] \in \{0, 1\}    \\
          %& M = m_1 \lor m_2 \lor ... \lor m_k, m_i \in \{l_1 \times w_1 \hspace{0.2em} rect,\hspace{0.2em} l_2 \times w_2 \hspace{0.2em} rect, ... \hspace{0.2em} \text{within     } O\}
    \end{aligned}
\end{equation}
where the second term is defined as \textit{\eotmetric}, computed as an expectation of attack success rate. In practice, we estimate this with a set of randomly sampled transforms. We observe that directly solving this objective is generally challenging, especially in the hard-label setting as we see in Section~\ref{sec:black-box}. Furthermore, the problem of finding an optimal solution with a constrained mask size is NP-complete, as we prove in Appendix~\ref{ap:mask_gen}. One approach to constructing such masks is to adopt an $\ell_0$ ``heuristic" (the $\ell_0$ norm itself is fundamentally non-differentiable, so straightforward adaptations of gradient descent or OPT-attack cannot be used). As an example, JSMA~\cite{papernot2016limitations} builds a mask by considering the pixels most ``relevant'' towards causing a misclassification and Carlini and Wagner's $\ell_0$ attack~\cite{carlini2017towards} reduces a mask by removing least ``relevant" pixels; both approaches require doing a large number of forward passes over the model (e.g., a pass for each pixel) to find most relevant or least pixel, since the problem is non-differentiable.  This forward-pass method works  in the white-box setting but as further discussed in Section~\ref{sec:black-box}, does not work in the hard-label setting where logits are not available; changing a pixel is unlikely to change the classification result, making it challenging to identify most relevant or least relevant pixels.

\subsection{General Algorithmic Pipeline}
We describe the general \alg solver framework with pseudocode in Algorithm~\ref{alg:general}. As the $\ell_0$ norm inherently makes~\eqref{eq:joint} difficult to solve directly, we alternate between reducing the mask size and improving the attack with the set of perturbable pixels. After initializing a perturbation and gradient, the framework iteratively selects pixels to remove, removes them, and attacks the remaining pixels. This process is repeated until a specified stopping criteria is met. %These steps inherently require that there exists some way to order pixelssuggest that the framework generally requires an algorithm to compute transform-robust  attacks and a way to order pixels.

The C\&W $\ell_0$ attack is a natural starting point for our framework as it alternates between removing the pixel with the least impact and attacking the remaining pixels with C\&W $\ell_2$ until it can no longer find an adversarial example (Algorithm~\ref{alg:cwl0}). Impact is measured by multiplying the gradient and the perturbation. We are able to adapt the C\&W $\ell_0$ attack to incorporate transform-robustness for our white-box \alg algorithm (Section~\ref{sec:white-box}).

We first consider simple adaptations of the C\&W $\ell_0$ attack, such as combining it with variations of OPT~\cite{cheng2018query} and EOT~\cite{athalye2017synthesizing}, but find that they fail to  find satisfactory hard-label physical attacks (Section~\ref{sec:baselines}).  We consider these adaptations  (listed in rows 3-5 of Table~\ref{tab:baselines}) as "baselines" for our algorithm.   This then motivates us to separate the optimization into a mask generation phase and perturbation boosting phase, which leads to our novel hard-label black-box \alg algorithm (Section~\ref{sec:bb_graphite}). Table~\ref{tab:comparisons} summarizes our algorithms and baselines. %The C\&W $\ell_0$ attack is a natural starting point for our framework (see Algorithm~\ref{alg:cwl0}), as it alternates between removing the pixel with the least impact (measured by multiplying the gradient and the perturbation) and then running an inner C\&W $\ell_2$ attack on the remaining pixels until the attack can no longer find an adversarial example. We will see in Section~\ref{sec:white-box} that we can adapt it for our white-box attack to automatically generate transform-robust attacks. However, we will show in Section~\ref{sec:baselines} that such adaptations fail to create satisfiable results in the hard-label setting, even with the addition of the digital hard-label OPT-attack~\cite{cheng2018query} ($\ell_0$ and OPT, $\ell_0$ and OPT w/ EoT, $\ell_0$ and boosting). This motivates us to separate out the optimization into a mask generation phase and perturbation boosting phase, which leads us to our hard-label solver in Sections~\ref{sec:start_graphite_explanation}-\ref{sec:boost}.

%  For ease of algorithmic notation in future sections, given a mask $M$ and perturbation $\delta$ the \eotmetric (i.e., the expectation term in~\eqref{eq:joint}) can be estimated by computing $n$ randomly
%  sampled transformations $t_i \sim T$: 

%  \begin{equation}\label{eq:survestimate}
%      S_{x,\target}(M,\delta) = \frac{ \sum\limits_{i=1}^{n} \mathbbm{1}\bigg[   F\bigg(t_i(x + M \cdot \mathbf{\delta})\bigg) = \target      \bigg]}{n}
%  \end{equation}

 \begin{algorithm}[t!]
\caption{General \alg Framework}\label{alg:general}
\begin{algorithmic}[1]
\Require Victim Image $x$, Target Image $x_{tar}$, Initial Mask $M_{init}$, Model $F$, Target Label $y_{tar}$
\Ensure Attacked Image $A$, Mask $M$, Perturbation $\delta$
\State $M \gets M_{init}$
\State $\delta, g \gets$ INIT\_PERT\_+\_GRAD($x, x_{tar}, M, F, y_{tar}$)
\While{not done}
    \State $S \gets$ SELECT\_PIXELS($x, x_{tar}, M, \delta, y_{tar}$, $g$)
    \State $M \gets$ REMOVE\_PIXELS($M, S$)
    \State $A, \delta, g \gets$ ATTACK($x, x_{tar}, M, \delta_{init}, F, y_{tar})$

\EndWhile
\State $A, \delta \gets$ Last Successful Attack
\end{algorithmic}
\end{algorithm}

%% file: tex/whitebox.tex
\section{White-box Automatic Physical Attacks}\label{sec:white-box}
We now create a white-box \alg algorithm as an instantiation of the general \alg framework (Algorithm~\ref{alg:general}). We do so by extending the Carlini-Wagner $\ell_0$ attack~\cite{carlini2017towards} to have high \eotmetric, creating an automatic, physical attack in the white-box setting.

%In this section, we will describe optimization problems for our approach, \alg. We want \alg to, without gradients, find a small candidate mask and perturbation within that masked area so that the resulting image is likely to be robust when physically realized. We also want \alg to be efficient and not have obvious target image artifacts. We will first describe a joint optimization problem for finding an optimal perturbation and mask, and then discuss a two-stage process that is more practical for the hard-label setting. The first stage finds an appropriate mask, and then the second stage optimizes the perturbation with RGF~\cite{nesterov2017random} in a reformulation that considers \eotmetric..

%As in Athalye \etal~\cite{athalye2017synthesizing}, we also rely on estimating \eotmetric of an image under transforms, i.e., the expectation of an image being predicted with a given label under physical transformations (this was termed EoT in Athalye \etal~\cite{athalye2017synthesizing}). Details on the specific transformations used are included in the supplementary material (\rf{fix this}).

\textbf{White-box Attack Algorithm.}
%\label{sec:supp_whitebox}
%Based on the general GRAPHITE framework, we first aim to create an automatic, physical attack in the white-box setting. We thus design an attack inspired by the Carlini-Wagner (C\&W) $\ell_0$ attack~\cite{carlini2017towards} modified to generate physical attacks. %The original C\&W $\ell_0$ algorithm, as presented in Algorithm~\ref{alg:cwl0}, fits into our framework. Specifically, the original algorithm greedily reduces the cardinality of the mask (allowed set of pixels that can be perturbed), alternating between pixel removal and running an inner C\&W $\ell_2$ attack until the attack can no longer find an adversarial example. The attack removes the pixel with the least impact, measured by multiplying the gradient and the perturbation together.
%determines impact with a heuristic: supposing that $x$ is the input image and $\delta$ is the perturbation, it removes the pixel $i = \argmin_{i}g_i \cdot \delta_i$, where $g_i$ is the gradient of the loss with respect to the $i^{th}$ pixel of the attacked image (i.e., $x + \delta$) and $\delta_i$ is the perturbation for that pixel from the attacked image. 
We extend the C\&W $\ell_0$ attack to add the notion of \eotmetric by replacing the inner C\&W $\ell_2$ attack with an EoT PGD attack. By using an inner attack that applies EoT~\cite{athalye2017synthesizing}, we are able to generate attacks that have higher transform-robustness. Then, replacing the C\&W $\ell_2$ minimization attack with PGD~\cite{madry2017towards} enables us to increase efficiency and align our attack more similarly to our later black-box implementation. With these changes, we now set the stopping criteria as our EoT PGD attack being unable to find an attack with at least $tr_{min}$ \eotmetric.

Additionally, for further efficiency and to encourage larger sticker patches (for printing convenience), we remove $z$ patches of pixels at a time, where each patch is a $p \times p$ square of pixels. To collect the list of patches, we stride by a step size of $s$.

\begin{algorithm}[t]
\caption{Original C\&W $\ell_0$ Attack}\label{alg:cwl0}
\begin{algorithmic}[1]
\Require Victim Image $x$, Model $F$, Initial Mask $M_{init}$, Target Label $y_{tar}$
\Ensure Attacked Image $A$, Mask $M$, Perturbation $\delta$
\State $M \gets M_{init}$
\State $A, \delta, g \gets$ C\&W\_$\ell_2$\_ATTACK($x, M, F, y_{tar}$)
\While{$F(A) = y_{tar}$}
    \State $S \gets$ $\{\argmin_i g_i \cdot \delta_i\}$
    \State $M_{\{Pixels \in S\}} \gets$ 0
    \State $A, \delta, g \gets$ C\&W\_$\ell_2$\_ATTACK($x, M, F, y_{tar}$)

\EndWhile
\State $A, \delta \gets$ Last Successful Attack
\end{algorithmic}
\end{algorithm}

 \begin{algorithm}[t]
\caption{White-box \alg Attack}\label{alg:wb}
\begin{algorithmic}[1]
\Require Victim Image $x$, Model $F$, Initial Mask $M_{init}$, Target Label $y_{tar}$
\Ensure Attacked Image $A$, Mask $M$, Perturbation $\delta$
\State $M \gets M_{init}$
\State $A, \delta, g \gets$ EoT\_PGD\_ATTACK($x, M, F, y_{tar}$)
\While{Transform-Robustness of A $> tr_{min}$}
    \State $S \gets$ {set of $z$ patches with smallest $g_i \cdot \delta_i$ values}
    \State $M_{\{Patches \in S\}} \gets$ 0
    \State $A, \delta, g \gets$ EoT\_PGD\_ATTACK($x, M, F, y_{tar}$)

\EndWhile
\State $A, \delta \gets$ Last Attack with \eotmetric $> tr_{min}$ if one was found, else first attack 
\end{algorithmic}
\end{algorithm}

%More specifically, we begin by initializing the mask (allowed set) of pixels to be all the pixels in the victim image. Then, we perform a round of \textit{mask reduction}, which comprises 2 steps: (1) Constructing an adversarial example with EoT~\cite{athalye2017synthesizing} and Projected Gradient Descent (PGD)~\cite{madry2017towards}, and (2) removing the pixels from the mask that are least relevant in causing a misclassification. We compute the average impact of the patches (in just the areas that overlap with the current mask) and remove the $z$ least impactful patches (a hyperparameter). For efficiency, each round begins with the previous round's perturbation and then applies a random start of size. This process iteratively repeats until the final round, in which step (1) can no longer find an adversarial example with \eotmetric greater than $S_t$ within a reasonable number of PGD iterations.

%At a high level, the algorithm greedily reduces the cardinality of the mask (allowed set of pixels that can be perturbed) until a \eotmetric threshold $S_t$ can no longer be achieved. Pixel relevancy is determined using a modification of the CW $L_0$ heuristic, which removes the pixel $i = \argmin_{i}g_i \cdot \delta_i$, where $g_i$ is the gradient of the loss with respect to the $i^{th}$ pixel, and $\delta_i$ is the perturbation for that pixel. For a detailed explanation of this heuristic, we refer readers to the original work. To adapt for the physical domain, we operate on patches of pixels instead of individual pixels. 

\begin{table*}
\centering
\caption{White-box comparisons between our white-box version of \alg (Algorithm~\ref{alg:wb}) and a Patch-PGD attack that tries the 4 corners and the center on 90 victim-target traffic sign pairs. While Patch-PGD can perform well on the right examples, \alg can much more reliably and quickly find successful attacks with less perturbation size. TR stands for \eotmetric. Gray columns report statistics over just the samples with $> 80\%$ TR.}\label{tab:wb_results}
\definecolor{Gray}{gray}{0.9}
\newcolumntype{a}{>{\columncolor{Gray}}c}
\begin{tabular}{@{}ccccaac@{}}
\toprule
\textbf{Method}                                                                        & \textbf{\begin{tabular}[c]{@{}c@{}}Avg. TR,\\ All\end{tabular}} & \textbf{\begin{tabular}[c]{@{}c@{}}Avg. Mask Size,\\ All\end{tabular}} & \textbf{\begin{tabular}[c]{@{}c@{}}Num. Samples\\ with \textgreater 80\% TR\end{tabular}} & \textbf{\begin{tabular}[c]{@{}c@{}}Avg. TR over\\ Samples with\\ \textgreater 80\% TR\end{tabular}} & \textbf{\begin{tabular}[c]{@{}c@{}}Avg. Mask Size over\\ Samples with\\ \textgreater 80\% TR\end{tabular}} & \textbf{Avg. Inferences} \\ \midrule
\textbf{GRAPHITE}                                                                      & 77.53\%                                                         & 8.82\%                                                                 & 84                                                                            & 80.60\%                                                                               & 6.35\%                                                                                       & 40,475.6                  \\ \midrule
\textbf{\begin{tabular}[c]{@{}c@{}}Patch-PGD\end{tabular}} & 10.1556\%                                                       & 6.88\%                                                                 & 4                                                                             & 95.25\%                                                                               & 6.88\%                                                                                       & 100,000                   \\ \bottomrule
\vspace{-0.2in}
\end{tabular}
\end{table*}

\textbf{Experimental Setup.}
To test our attack, we run targeted attacks for all victim-target pairs in a varied, 10 class subset of GTSRB: \texttt{Stop, Speed Limit 30 km / hr, Speed Limit 80 km / hr, Pedestrians, Turn Left Ahead, Yield, Caution, Roundabout, End of Overtaking Limit, Do Not Enter}. Foreshadowing our black-box attack, where we want to show dataset independence, we use images outside of the GTSRB dataset. Since we had a real-world \texttt{Stop Sign} available in our lab, we used a photo we took of that physical sign. For all other GTSRB classes, we attack traffic-sign images downloaded from the Internet. The images are $244 \times 244$. We set the minimum transform-robustness threshold $tr_{min}$ to $80\%$. Additional details on hyperparameters are included in Appendix~\ref{ap:wb_hyper}.

\textbf{Patch-PGD.}
For comparisons sake, we also compared against the popular square patch attack that simply runs Patch-PGD~\cite{chiang2020certified,chong2021patchguard} (with EoT) attack over a square patch of $64 \times 64$ pixels (about 6.88\% of the image).  Rather than test every possible location, which is expensive, we test the four corners (as in the weaker adversary in~\cite{chiang2020certified}) and the center. We ran it with 100 transforms, 200 steps, and a step size of 4 / 255.

\textbf{Results.}
Table~\ref{tab:wb_results} presents the results from our white-box algorithm and the Patch-PGD baseline on all 90 victim-target pairs of the GTSRB subset. We also report results on the images with $>$ 80\% \eotmetric (the gray columns) to characterize the white-box \alg attacks that reached the $tr_{min}$ threshold. 

Our approach finds patches with higher \eotmetric more reliably and efficiently than choosing static squares, even when using pre-selected patch locations. Patch-PGD finds attacks with low \eotmetric (10.1556\% on average, only 4 samples with $> 80\%$) and a high number of inferences (100k). In contrast, our white-box \alg algorithm averages 77.53\% \eotmetric, including 84 samples with $> 80\%$. It also uses fewer inferences and achieves a smaller mask size over the successful samples (6.35\% for \alg vs. 6.88\% for Patch-PGD. Finally, our white-box algorithm does not require a user to specify the shape or size of the mask or its location. Example images of these attacks are included in Appendix~\ref{ap:wb_hyper}. 

%%%%%%%%%%%%%%5 OLD white box results

%% file: tex/blackbox.tex
\section{Black-box (Hard-label) Automatic Physical Attacks}\label{sec:black-box}

% \begin{figure*}
%     \centering
%     \includegraphics[width=0.8\textwidth]{figs/pipeline euro.pdf}

% \caption{Figure showing the algorithm flow for a targeted attack on a \texttt{Stop sign} to \texttt{Speed Limit 30 km / hr sign}. Multiple transformations are sampled and used at each stage.} \label{fig:flow}
% \end{figure*}

%%%%%%%%%% OLD Black box algo

We now create a new hard-label \alg algorithm as an instantiation of the general \alg framework (Algorithm~\ref{alg:general}) to generate automatic physical attacks in the hard-label setting. Our white-box algorithm (Section~\ref{sec:white-box}) cannot be trivially extended here, as our prior selection of pixels for removal required gradient access. 

%\rf{because CW doesn't work in black box, we cannot solve joint optimization directly. so we do two stage decoupling}

We first attempt to leverage the C\&W $\ell_0$ algorithm in combination with existing digital black-box attacks, with three new ``baseline'' black-box instantiations of \alg (Section~\ref{sec:baselines}). However, we find  these ``baselines" to be unsatisfactory (see the results in Section~\ref{sec:baselines}).

To address these limitations, we split the joint optimization problem in~\ref{eq:joint} into a two-step optimization instead (Section~\ref{sec:start_graphite_explanation}). Within this two-step process, we then (a) present a novel mask generation algorithm (Section~\ref{sec:mask_gen}) for the pixel selection and removal procedures (lines 4 and 5 of Algorithm~\ref{alg:general}) and (b) adopt Randomized Gradient Free (RGF) optimization of \eotmetric itself as the attack procedure (line 6 of Algorithm~\ref{alg:general}), referred to as ``perturbation boosting'' (Section~\ref{sec:boost}).

%This section is organized as follows: we begin by again leveraging the C\&W $\ell_0$ algorithm, now in combination with existing digital black-box attacks, to present three ``baseline'' black-box instantiations of \alg in Section~\ref{sec:baselines}. Upon finding that these ``baselines'' are insufficient, rather than solving the joint optimization problem directly we decouple the optimization into a two-step optimization (Section~\ref{sec:start_graphite_explanation}). We then (a) present a novel algorithm for the pixel selection and removal procedures (lines 4 and 5 of Algorithm~\ref{alg:general}), referred to as ``mask generation'', and (b) adopt Randomized Gradient Free (RGF) optimization of \eotmetric itself as the attack procedure (line 6 of Algorithm~\ref{alg:general}), referred to as ``perturbation boosting'' in Sections~\ref{sec:mask_gen}-\ref{sec:boost}.

% for mask generation Eac each comprising a modification to our white-box algorithm by replacing the inner attack (line 6 of Algorithms~\ref{alg:general},~\ref{alg:wb}) with a suitable black-box attack that leverages estimated gradients. We then attempt to instead approximate this problem, by formulating the attack as a two-step optimization problem that separates mask generation and perturbation boosting. Finally, we present algorithms for hard-label mask generation and perturbation boosting and analyze the suitability of our perturbation boosting formulation.

\subsection{Baselines}\label{sec:baselines}
These baselines modify the C\&W $\ell_0$ paradigm of alternating pixel removal and attacking with each baseline differing only in terms of the inner attack choice. For our first two baselines, we adopt a vanilla and EoT version of the digital black-box OPT-attack~\cite{cheng2018query}. OPT-attack is able to craft attacks in the black-box setting and provide estimated gradients that can be used to order pixels for pixel selection and removal. For our third baseline, we deviate from OPT-attack and instead adopt perturbation boosting --- our RGF optimization attack that leverages \eotmetric itself as the objective function.

\subsubsection{Setup}
To evaluate our baselines in this section, we run targeted attacks for all 90 victim-target pairs in the 10 class subset of GTSRB used in Section~\ref{sec:white-box}. All hyper-parameters are identical to those used in our white-box experiments, and performed on 32x32 images. We also initialize the perturbation direction $\theta$ as the difference between the target and victim images.

\begin{table*}
    \centering
    % \resizebox{\columnwidth}{!}{
    \caption{Quantitative results for hard-label baseline attacks. We report the average and standard deviation of \eotmetric, \# of pixels out of 32 x 32 = 1024 in the mask (i.e., $\ell_0$ distance), and number of queries issued to the model. Statistics are presented only for successful attacks (i.e., attacks that are able to craft an adversarial example).}
    \begin{tabular}{ccccc}
    \toprule
         \bf Baseline & \bf \# Successes &  \bf \EOTmetric (\%) & \bf Queries & \bf Mask Size\\
         \midrule
         $\ell_0$ and OPT & 90/90 & 1.52 $\pm$ 4.26 & 919.82k $\pm$ 605.24k  & 244.36 $\pm$ 307.11 \\
         $\ell_0$ and OPT with EoT & 81/90 & 80.01 $\pm$ 0.11 & 1958.67k $\pm$ 1317.49k & 601.90 $\pm$ 190.91 \\
         $\ell_0$ and Boosting & 81/90 & 97.89 $\pm$ 3.61 & 5.50k $\pm$ 9.17k & 990.72 $\pm$ 56.78 \\
         \bottomrule
    \end{tabular}    
    % }

    \label{tab:baselines}
\vspace{-0.1in}
\end{table*}

\begin{table}
    \centering
    \caption{Samples from our three hard-label baseline attacks. Shown are targeted attacks from a \texttt{Stop sign} to \texttt{Speed Limit 30 km / hr sign}. The leftmost image suffers from extremely low transform-robustness. The middle and rightmost images exhibit higher transform-robustness at the cost of extreme visible artifacts from the target image.}
    \begin{tabular}{c|c|c}
    \toprule
        \textbf{$\ell_0$ and OPT} & \textbf{$\ell_0$ and OPT w/ EoT} & \textbf{$\ell_0$ and Boosting} \\
        \midrule
        \includegraphics[width=0.25\columnwidth]{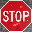} & \includegraphics[width=0.25\columnwidth]{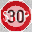} & \includegraphics[width=0.25\columnwidth]{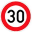} \\
     \bottomrule
    \end{tabular}
    \label{tab:baseline_imgs}
    \vspace{-0.1in}
\end{table}
\subsubsection{\texorpdfstring{$\ell_0$}{l0} and OPT baseline}\label{subsec:l0opt}
This baseline employs the vanilla OPT-attack~\cite{cheng2018query} as its choice of attack algorithm. OPT-attack crafts an adversarial example using a distance-based objective:
\begin{equation}\label{eqn:opt_attack_baseline}
        \argmin_\theta g(\theta)
\end{equation}

where the objective $g(\theta)$ is the distance to the decision boundary, formally defined as: 
\begin{equation}\label{eq:g}
            \argmin\limits_{\mu > 0} \; \mu \;\;s.t.\;\; \left(F\left(x + \mu \frac{\mathbf{\theta}}{\vert\vert\mathbf{\theta}\vert\vert_2}\right) = \target\right)
\end{equation}

and where $\theta$ is the direction and $\mu$ is the distance to the nearest adversarial example $x^*$ in that direction. We consider an attack a success so long as it elicits misclassification, regardless of \eotmetric (as transforms have not been incorporated in any manner).

%This attack is suitable for \alg --- it returns both an adversarial example, and estimated gradients for~\eqref{eqn:opt_attack_baseline} which can be used for pixel selection/removal. 
%We consider an attack a success so long as it elicits misclassification, regardless of \eotmetric (as transforms have not been incorporated in any manner).

\textbf{Results.} We find that OPT-attack is generally unable to find attacks with any significant transform-robustness (Table~\ref{tab:baselines}, Table~\ref{tab:baseline_imgs}). In particular, the average \eotmetric is only $\sim 1.5\%$. This is likely an artifact of~\eqref{eqn:opt_attack_baseline} not incorporating transforms. OPT-attack also results in a significant number of queries ($\sim 920$k) because of its intense decision boundary-searching procedures. The final mask size is the only aspect of this baseline that is modestly respectable, with a perturbation that covers $\sim 24\%$ of the input image.

\subsubsection{\texorpdfstring{$\ell_0$}{l0} and OPT with EoT baseline}\label{subsec:l0opteot} This baseline employs an EoT version of OPT-attack~\cite{cheng2018query} as its choice of attack algorithm, to alleviate the low \eotmetric of the first baseline. We first define a wrapper function $W(x)$:
% Most concerning of the poor results from our first baseline was the extremely low transform-robustness of attacked images. As such, our next baseline aims to directly address this shortcoming. Specifically, this baseline introduces EoT into OPT-attack, by averaging the computed gradients across several randomly transformed versions of the input image and finding the minimum distance with an acceptable \eotmetric. We first define a wrapper function $W(x)$:
\begin{equation}
  \begin{aligned}
  W(x) = \begin{cases}
     y & \text{if}\hspace{0.2em} \E_{t \sim T} \bigg[ F\bigg(t(x)\bigg) = \target \bigg] \geq 80\%\\
      -1 & \text{otherwise} 
   \end{cases}
  \end{aligned}
\end{equation}

Then, with this wrapper function, we modify $g(\theta)$ to the following:
\begin{equation}\label{eqn:opt_attack_eot_baseline}
    \begin{split}
            \argmin\limits_{\mu > 0} \; \mu \;\;s.t.\;\; \left(W\left(x + \mu \frac{\mathbf{\theta}}{\vert\vert\mathbf{\theta}\vert\vert_2}\right) = \target\right)
    \end{split}
\end{equation}

We now consider an attack a success only if it exhibits a \eotmetric greater than the $tr_{min}$ threshold, as this baseline directly incorporates transforms via EoT.

%This attack is again suitable for \alg, for the same reasons as before. However, we now consider an attack a success only if it exhibits an \eotmetric greater than the $tr_{min}$ threshold, as this baseline directly incorporates transforms via EoT.

% Given the objective in~\eqref{eqn:opt_attack_eot_baseline}, we simply perform a round of mask reduction by computing the averaged gradients of this objective with respect to each pixel. This process iteratively repeats until the  final  round,  in  which  OPT  attack  can  no  longer  find an adversarial example with \eotmetric greater than $S_t = 80\%$ with the current mask (within a reasonable number of iterations). We note that if OPT attack cannot find an example with \eotmetric greater than $S_t$ within at least one round (i.e., with the full mask), then we consider the attack a failure.

\textbf{Results.} We find that wrapping existing hard-label attacks with EoT is an inefficient approach to generating attacks with high transform-robustness (Table~\ref{tab:baselines}, Table~\ref{tab:baseline_imgs}). While, as expected, the introduction of EoT significantly raises average \eotmetric to $\sim 80\%$, it comes at the cost of doubling in queries $\sim 1,959$k.

Further, we find this approach yields attacks with unacceptably large target image artifacts. The average mask size covers nearly $59\%$ of the input image. We also observe that even for the successful attacks, OPT attack is unable to escape the initial perturbation direction (i.e., the target image itself). This results in attack examples with very visible artifacts from the target. 

\subsubsection{\texorpdfstring{$\ell_0$}{l0} and Boosting Baseline}\label{subsec:l0boost}
This baseline employs perturbation boosting as its choice of attack algorithm, i.e., an RGF optimization attack that leverages \eotmetric itself as the objective function. This choice is motivated by the intuition that directly optimizing for our desired metric (instead of a proxy objective function provided by existing black-box attacks) is likely a more straightforward optimization problem, which is likely to result in fewer queries and a smaller mask, while still achieving high \eotmetric. For further details regarding the details and suitability of perturbation boosting for \alg, see Section~\ref{sec:boost}, where it becomes a part of our final, recommended instantiation algorithm for \alg in the black-box setting.

% Our observations from the $\ell_0$ and OPT with EoT baseline indicated that we can obtain attacks with sufficient \eotmetric to be a  potential real-world concern, at the cost of queries, unnaceptably large masks, and visible artifacts from the target image. This suggests that OPT attack's distance based formulation is a sub-optimal solution to generating such attacks in physical hard-label settings. A natural solution to address this sub-optimality is to replace OPT-attack with an attack that \textit{directly maximizes \eotmetric}. In other words, this baseline replaces the OPT and EoT objective in ~\eqref{eqn:opt_attack_eot_baseline} with the transform-robustness objective itself (see \eqref{eq:survestimate})\rf{fix}. For more details on this attack, see Section~\ref{sec:boost}, as this component becomes part of the final \alg pipeline (from now on, \alg refers to the algorithm for hard-label blackbox setting).

\textbf{Results.} We find that this baseline yields transform-robust attacks with unacceptably large masks (Table~\ref{tab:baselines}, Table~\ref{tab:baseline_imgs}). While this baseline achieves a near optimal $\sim 98\%$ \eotmetric as expected (since we are now directly maximizing \eotmetric), the mask sizes cover nearly $97\%$ of the input image. 

Close inspection reveals that this is because the estimated RGF optimization gradients eventually become zero at some iteration (i.e., the area around the example appears to be flat, thereby obstructing sampling-based based techniques like RGF). Zero gradients prevent the algorithm from creating a reliable importance ordering of pixels, thereby ending the attack as pixel selection/removal cannot proceed. This also results in several artifacts from the target image, as 1-2 rounds are not enough to create a sufficiently small mask or produce a more obscure perturbation. These early exits also explain how this baseline is able to operate with significantly fewer queries than previous baselines ($\sim 5.5$k).

\subsection{Black-box GRAPHITE Algorithm}\label{sec:bb_graphite}
Our black-box baseline instantiations in Section~\ref{sec:baselines} failed to craft acceptable automatic physical attacks. However, these failures provided key insights towards developing automatic physical attacks: (a) creating high \eotmetric examples requires direct incorporation of transforms (EoT) into the attack, (b) na\"{i}ve application of EoT to existing black-box attacks is an indirect and query-inefficient approach to doing so, and (c) reliance upon leveraging gradient-based pixel ordering can fail in the black-box setting, thus leading to large masks. 

We instead solve a \emph{two-step optimization problem} that separates out mask generation and boosting (Section~\ref{sec:start_graphite_explanation}). While the C\&W $\ell_0$ framework of alternating between pixel removal and attacking is able to capture a joint optimization in the white-box setting, our baselines showed that jointly optimizing the mask and the perturbation in the hard-label setting is difficult without gradients. Thus, we instead focus on generating a viable, small mask first before focusing on perturbation boosting. We then present algorithms for solving these optimization problems (Sections~\ref{sec:mask_gen}-\ref{sec:boost}).

\subsubsection{Two-Step Optimization}\label{sec:start_graphite_explanation}

% As we saw in Section~\ref{sec:baselines}, simple extensions of OPT, EoT, and C\&W $\ell_0$ attacks fail to generate good automatic, physical, hard-label attacks. This largely comes down to an incompatibility with the way the hard-label gradients were obtained and the application of EoT. In the $\ell_0$ and OPT with EoT baseline, the hard-label initialization and neighborhood sampling creates a dependence on starting at a good point on a smooth objective function. Unfortunately, we are unable to reliably get out of the initial perturbation direction and get stuck with examples that can largely be found by interpolating between victim and target images. In the $\ell_0$ and boosting baseline, we found that the hard-label gradients failed to reliably create an ordering of pixels based on relevancy.

We now present a two-step optimization problem that first solves for an optimal mask given an initial perturbation and then second solves for the optimal perturbation given the discovered mask. This two-step problem can be viewed as heuristically solving the general framework presented in Algorithm~\ref{alg:general} with a single iteration, with mask generation corresponding to pixel selection and removal and perturbation boosting becoming the internal attack. This solution then forms the basis for our primary, recommended hard-label instantiation of \alg, as shown in Algorithm~\ref{alg:bb}.

\textbf{Step \#1: Mask Generation.}
We first find an optimized mask that, when filled in with the target image $x_{tar}$, results in at least a specified level of transform-robustness, $tr_{lo}$, a hyperparameter. We constrain the unmasked perturbation to be fixed at $\delta_{tar} = x_{tar} - x$ and solve the optimization objective from~\eqref{eq:joint} under that constraint:
\begin{equation}\label{eq:mask}
    \begin{aligned}
        \argmin_{M} \; & \lambda\cdot \vert\vert M\vert\vert_0 %\\%+ \lambda_2\cdot h(M) \\
        - \E_{t \sim T} \bigg[ F\bigg(t(x + M \cdot \delta_{tar})\bigg) = \target \bigg]\\
        \textrm{s.t.} \; &  \E_{t \sim T} \bigg[ F\bigg(t(x + M \cdot \delta_{tar})\bigg) = \target \bigg] \ge tr_{lo}     
          %& M = m_1 \lor m_2 \lor ... \lor m_k, m_i \in \{l_1 \times w_1 \hspace{0.2em} rect,\hspace{0.2em} l_2 \times w_2 \hspace{0.2em} rect, ... \hspace{0.2em} \text{within     } O\}
    \end{aligned}
\end{equation}
We show how to solve this formulation heuristically using an algorithm described in Section~\ref{sec:mask_gen}. Note that $\delta_{tar}$ only applies within the resulting mask $M$.

\textbf{Step \#2: Perturbation Boosting.}
Equipped with the mask $M$ found by the Mask Generation Optimization Problem in Step \#1, we then aim to maximize \eotmetric for the given mask by changing the perturbation $\delta$ within the masked region:

\begin{equation}\label{eq:boost_full}
    \begin{aligned}
        \argmax_{\delta} \quad & \E_{t \sim T} \bigg[ F\bigg(t(x + M \cdot \delta)\bigg) = \target  \bigg]\\
       % \textrm{s.t.} \quad & M_g[i] \in \{0, 1\}\\
    \end{aligned}
\end{equation}

 \begin{algorithm}[t]
\caption{Black-box \alg Attack}\label{alg:bb}
\begin{algorithmic}[1]
\Require Victim Image $x$, Target Image $x_{tar}$, Model $F$, Initial Mask $M_{init}$, Target Label $y_{tar}$
\Ensure Attacked Image $A$, Mask $M$, Perturbation $\delta$
\State $M \gets M_{init}$
\State $\delta \gets M \cdot (x_{tar} - x)$
\For{One Iteration}
    \State // Select and Remove Patches (Mask Generation)
    \State $\mathcal{P} \gets$ HEATMAP\_ESTIMATION($M, \delta, F$)
    \State $M, \mathcal{P} \gets$ COARSE\_REDUCTION($M, \delta, F, P$)
    \State $M \gets$ FINE\_REDUCTION($M, \delta, F, P$)
    \State
    \State // Attack (Perturbation Boosting)
    \State $A, \delta \gets$ BOOST($x, \delta, M, F, y_{tar}$)

\EndFor
\end{algorithmic}
\end{algorithm}
By using \eotmetric as a measure of physical robustness, we can leverage this function as an optimization goal we can pursue even in the hard-label case (described in Section~\ref{sec:boost}). In Section~\ref{sec:boost}, we also provide empirical evidence that this  reformulation based on \eotmetric is approximately continuous. We do this by showing that the objective has relatively low Lipschitz constants even when approximating the expectation in~\eqref{eq:boost_full} by averaging over $n$ transformations. We additionally refer to this process as \emph{boosting} the \eotmetric, as it aims to make \eotmetric as high as possible.

% \subsection{\alg Framework Fundamentals}
% We propose solving this equation with the following general framework:
% \begin{enumerate}
%     \item Start with a full mask over the perturbable area
%     \item Estimate patch significance
%     \item Remove some number of patches
%     \item Boost the transform-robustness of the new partial mask
%     \item GOTO 2) or exit
% \end{enumerate}

% In this work, we mainly focus on one particular implementation of this framework that completes the above process once.

\subsubsection{Mask Generation Algorithm}\label{sec:mask_gen}
For the first step of our hard-label \alg algorithm, we need to find a candidate mask that has good initial \eotmetric and small size. Boosting can take care of further improving the \eotmetric as long as it samples enough variance at nearby points to make useful gradient estimates with RGF. We first set the perturbation to the target image's pixels within the mask. In other words, the initial attack is $x + M\cdot (x_{tar} - x)$.

\begin{algorithm}[t]
% \setstretch{1.05}
    \caption{Mask Generation. Let $TR(M, \delta)$ be the estimate of transform-robustness for a mask $M$ and perturbation $\delta$, and $\lambda_1$ be a hyperparameter, let $\mathcal{P}$ be the set of all patches, let $M$ be the mask covering the whole object, and let $\delta_{tar} = x_{tar} - x$.}\label{alg:mask_gen}
\begin{algorithmic}[1]
\Require Victim Image $x$, Target Image $x_{tar}$, Target Label $y_{tar}$, Optional Max Mask Size $m_{max}$
\Ensure Mask $M$ with lowest mask score $J(M, \delta, tr_{lo})$
\Function{$J$}{$M$, $\delta$, $tr_{lo}$}
  \If{$TR(M, \delta) < tr_{lo}$}
  \State \textbf{return} $\infty$
  \EndIf
  \State \textbf{return} $\lambda_1 \cdot \vert\vert M\vert\vert_0 + \big(1 - TR\big(M, x_{tar} - x\big)\big)$
\EndFunction\\

%\State $\mathcal{P} \gets $ Set of all patches
%\State $M \gets $ Mask covering the whole object
%\State $\delta_{tar} \gets x_{tar} - x$\\
\State{// HEATMAP ESTIMATION}
\For{$\rho$ in $\mathcal{P}$}
    \State $tr_\rho \gets TR(M - \rho, \delta_{tar})$

\EndFor

\State Sort $\mathcal{P}$ from highest $tr_\rho$ to lowest\\

% \begin{spacing}{2}
\State // COARSE REDUCTION
\State $i \gets$ Binary search for highest index such that
\vspace{0.05in}
$TR\big((\bigcup_{i=pivot}^{|\mathcal{P}|} \mathcal{P}_{i}), \delta_{tar}\big) >= tr_{hi}$
\State $L_c \gets$ \{$\mathcal{P}[i], \mathcal{P}[i+1], ..., \mathcal{P}[\vert\mathcal{P}\vert - 1]$\}
\vspace{0.05in}
\State $M \gets$ $(\bigcup_{i=pivot}^{|\mathcal{P}|} \mathcal{P}_{i}$)\\
% \end{spacing}

\State // FINE REDUCTION
\For{$\rho$ in $L_c$}
    \If{$J(M - \rho, \delta_{tar}, tr_{lo}) < J(M, \delta_{tar}, tr_{lo})$}
        \State $M \gets M - \rho$
        \If{$m_{max}$ is provided and $\vert\vert M\vert\vert_0 < m_{max}$}
            \State \textbf{break}
        \EndIf
    \EndIf
\EndFor
\If{$m_{max}$ is provided and $\vert\vert M\vert\vert_0 > m_{max}$}
            \State Increase $\lambda_1$; Goto 24
        \EndIf
        
\State \textbf{return} $M$

\end{algorithmic}
\end{algorithm}

%We generate masks using a three stage process: 1) Heatmap Estimation, 2) Coarse-grained Mask Reduction, 3) Fine-grained Mask Reduction.

We generate masks using a three-stage process:

\begin{enumerate}
    \item  Heatmap Estimation
    \item Coarse-grained Mask Reduction
    \item  Fine-grained Mask Reduction
\end{enumerate}

The algorithm steps are shown in Algorithm~\ref{alg:mask_gen}. We first initialize the mask to the entire victim object, which overlays the target object over the victim. We then choose a group of pixels called a {\em patch} as a configurable input into the algorithm. We then collect all patches of a given shape (e.g. $4 \times 4$ squares) that overlap with the victim object. These patches serve two purposes:
(1) it helps us estimate a ``heatmap" (i.e., which regions of pixels contribute more to \eotmetric) and (2) it is used to remove groups of pixels from the mask to
reduce its size as the algorithm optimizes the mask objective function~\eqref{eq:mask}. The shape size such as $4 \times 4$ square  (rather than a single pixel) helps ensure that the attack image is not too pixelated and thus more likely to be implementable as a physical perturbation such as via stickers.

This process can be visualized by imagining a thin overlay of the target being placed on top of the victim, and then slowly cutting regions out of the overlay, exposing the victim object underneath.

\textbf{Heatmap Estimation.}
We generate a ``heatmap" (similar to a saliency map) over the victim image to begin pixel selection and removal. This could be any process that generates an ordering of pixel patches. We mainly focus on one strategy that we call a ``target-based heatmap" strategy, which orders patches by the \eotmetric with respect to the target.

Specifically, for each patch $p$ that overlaps with the victim object, we measure the \eotmetric of the mask $M$ and perturbation $\delta_{tar}$, where $M$ includes every pixel of the victim object except for $p$. In other words, we take the original target image overlay, cut out $p$, and measure the resulting \eotmetric. If the \eotmetric drops a lot relative to the \eotmetric with the patch after removing $p$, those target pixels are important to causing a target prediction. This enables us to identify candidate regions to remove from the mask. We output the sorted list of patches in decreasing order of transform-robustness.

The size of the patches matters in the hard-label setting. If we pick just a single pixel as our patch, the \eotmetric would likely be the same with or without it, making the heatmap useless. If we make the patch too large, we lose the ability to make more general mask shapes. For 32$\times$32 noise, we empirically found that a 4$\times$4 square patch worked well.

The ``target-based heatmap" approach can successfully evaluate patches with varying degrees of \eotmetric, leading to a useful ordering. One could imagine a ``victim-based heatmap" where we add small sections of the target onto the victim, calculating the \eotmetric with respect to the victim, but we found this approach to be sub-optimal. Our intuition is that this is a less direct measure in the targeted case - classifications of the target versus any other non-victim labels are treated the same. Note that in a ``victim-based" setting we cannot practically compute \eotmetric with respect to the target since the target signal is typically too small to get anything other than $\approx 0\%$ \eotmetric. We try a random heatmap strategy in Section~\ref{sec:extra_results}, where we simply order the patches randomly and pass the list on to fine-grained reduction.

\textbf{Coarse-grained Mask Reduction.}
Using the sorted list of patches from heatmap estimation,
we begin reducing the size of the mask. 
As an optimization to save queries, we first do a coarse-grained reduction that binary searches for a pivot in the patch list. We find the point $pivot$ in the ordered list of patches $\mathcal{P}$ such that the bitwise unions of patches from that point ($\bigcup_{i=pivot}^{|\mathcal{P}|} \mathcal{P}_{i}$) yields a mask of \eotmetric $\geq tr_{hi}$, where $tr_{hi}$ is a hyper-parameter specifying a high \eotmetric threshold that must be reached. If $tr_{hi}$ cannot be reached, we simply include all patches. Coarse-grained reduction outputs $L_c$, the list of patches from $pivot$ to the end.

\textbf{Fine-grained Mask Reduction.}
We then use a greedy fine-grained reduction algorithm to further improve the objective function of~\eqref{eq:mask}. Fine-grained reduction takes $L_c$ and evaluates each patch in sorted order. It removes the patch if that improves the objective function, keeping it otherwise. Empirically, we found that it was important to add the restriction that \eotmetric does not cross below some minimum threshold, $tr_{lo}$ to ensure success in boosting. The end result is the final mask defined by the union of the retained elements of $L_c$. Because the objective function rewards small masks and high \eotmetric, the final mask balances both well.

We can optionally specify to \alg that we desire a maximum mask size $m_{max}$. If this option is activated, we can simply break early once we drop below $m_{max}$. If the first iteration fails to get below $m_{max}$, then the value of $\lambda$ (the weight of the mask size term in the joint optimization problem in~\eqref{eq:joint}) can be increased until an iteration gets below $m_{max}$.

Let $n$ be the number of transforms used for estimating \eotmetric. Heatmap estimation uses $n \cdot |\mathcal{P}|$ queries, coarse-grained reduction uses $n \cdot \log{|\mathcal{P}|}$ queries, and fine-grained reduction uses $n \cdot |L_c|$ queries.

\subsubsection{Perturbation Boosting Algorithm.}\label{sec:boost}
Given a resulting image $x$ and a mask $M$ from the previous stage, \eotmetric boosting, or simply {\em boosting}, aims
to find the perturbation $\mathbf{\delta}$ that maximizes transform-robustness.  We propose a \emph{\eotmetric-based} reformulation to use with RGF~\cite{ghadimi2013stochastic,nesterov2017random} to find the perturbation within the mask $M$ with maximum \eotmetric, which works as follows. 

%Unlike Cheng et al.~\cite{cheng2018query}, we initialize $\mathbf{\theta}$ to be the difference between some given target class image $\mathbf{e}$ and the starting image $\bm{x}$ instead of taking the best such difference between the starting image $\bm{x}$ and 1000 training set target class images. This removes the dependency on knowing the training set. As before, we use a mask to restrict the perturbation to certain areas of the original image and only generate noise within the mask.

%It is important to note that when incorporating masks of the scale of RP${}_2$~\cite{roadsigns17,yolo} the initialization patch does not trivially yield an optimal answer, despite it turning parts of the victim image into an example target image. The fact that these patches do not survive as well as nearby perturbations that \alg can find makes it possible to optimize over this objective after initialization. 

The perturbation $\mathbf{\delta}$ is first initialized  to $M \cdot (x_{tar} - x)$ and then we proceed to maximize the probability that a perturbation remains robust to physical-world transforms with the RGF~\cite{ghadimi2013stochastic,nesterov2017random} method for gradient estimation using $q$ random samples for each gradient estimation. Explicitly, let $u$ be random Gaussian unit vectors within the allowable range of the mask and let $\beta$ be a nonzero smoothing parameter. Then, we set $\delta$ to $\delta - \eta \cdot \hat{g}$ where $\eta$ is the step size and $\hat{g}$ is the gradient, calculated as follows:

\begin{equation}
    \hat{g} = \frac{TR(M, \delta + \beta u) - TR(M, \delta)}{\beta} \cdot u
\end{equation}

where $TR(M, \delta)$ refers to the estimated \eotmetric for a mask $M$ and a perturbation $\delta$.

% \rf{fix para} Our \eotmetric-based reformulation has two key benefits. First, we find experimentally that it leaves fewer obvious artifacts of the target image (see Table~\ref{tab:digital_baselines}). Secondly, it is also more query-efficient than OPT-attack as it can measure \eotmetric directly. RGF requires frequent computations of the objective, which is \eotmetric in our case and boundary distance in OPT-attack. Calculating the boundary distance in OPT-attack requires performing query-intensive binary search operations while we do not.

Finally, we introduce the notion of a query budget which limits the number of queries used in a particular stage of the algorithm. These parameters can be tuned to emphasize better mask generation or better boosting without a budget of queries if we want to limit the number of attack queries.
%To ensure query efficiency, we also impose a query budget $b$ on the boosting stage.

\textbf{Suitability of Transform-Robustness Reformulation.}
To demonstrate that our reformulation based on \eotmetric is approximately continuous for successful RGF optimization, we show empirically that our objective has a low local Lipschitz constant. We execute boosting on attack examples from \texttt{Stop sign} to \texttt{Speed Limit 30 km/hr}, \texttt{Stop sign} to \texttt{Pedestrians}, and \texttt{Stop sign} to \texttt{Turn Left Ahead} with $n = 1000$ transformations and approximate the local Lipschitz constant every time we compute $TR(M, \delta + \beta \cdot u)$. The approximate local Lipschitz constant is given by $\frac{\vert TR(M, \delta + \beta \cdot u) - TR(\delta)\vert}{\vert\vert\beta\cdot u\vert\vert}$. We found that the maximum observed local Lipschitz constant was 0.056. We include a histogram of observed local Lipschitz constants in Fig.~\ref{fig:lipschitz}.%, which shows the majority of these values are very low. 

%% file: tex/experiments.tex
%------------------------------------------------------------------------
\section{Experiments}\label{sec:experiments}
In this section, we demonstrate the digital and physical viability of our black-box \alg instantiation presented in Algorithm~\ref{alg:bb} and Section~\ref{sec:bb_graphite}, henceforth referred to as simply \alg unless otherwise specified. We first report digital transform-robustness results on GTSRB~\cite{stallkamp2012man} and CIFAR-10~\cite{krizhevsky2009learning}. Then, we print stickers on a smaller subset of attacks, and conduct real-world field tests on physical traffic signs to show that our results carry over to the real-world. Finally, we explore our pipeline's ability to adapt to attacker priorities/constraints, allowing for a trade-off between \eotmetric, mask size, and query budget.% We also compare against digital OPT-attack and a simple OPT+EoT extension attack as baselines.

%%%%%%%%%%%%%%%%%%%%%%%%%%%%%%% OLD LIP LOCATION

\begin{figure}
    \centering
    \includegraphics[width=0.85\linewidth]{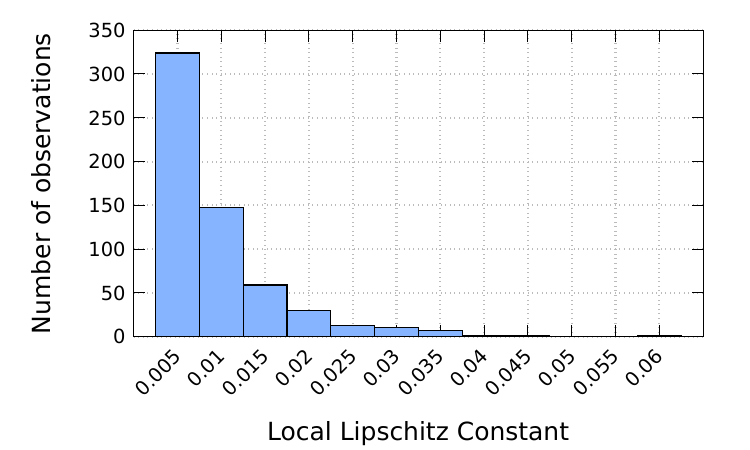}
    \caption{A histogram of our approximate local Lipschitz constant observations on \eotmetric. This graph shows that \eotmetric does not vary too much per amount of change with a sufficient number of transformations.}
    \label{fig:lipschitz}
    \vspace{-0.1in}
\end{figure}

%\input{tex/physicalattacks}

%-------------------------------------------------------------------------
\subsection{Experimental Setup}\label{sec:setup}
This section discusses our experimental setup for running \alg. \ifthenelse{\arxiv = 1}{}{
      We will publicly release our code and data prior to paper publication.}

\textbf{Datasets and Classifiers.}
We use the classifier from RP${}_2$~\cite{roadsigns17,yadav} trained on an augmented GTSRB dataset~\cite{stallkamp2012man}. As with RP${}_2$~\cite{roadsigns17}, we replace the German \texttt{Stop signs} with U.S. \texttt{Stop signs} from the LISA dataset~\cite{mogelmose2012vision}. As a validation set, we take out the last 10\% from each class in the training set. We augment the dataset with random rotation, translation, and shear, following Eykholt \etal~\cite{roadsigns17}. Our network, GTSRB-Net, has a 97.656\% test set accuracy.

\begin{table}
\centering
\caption{Hyperparameters for black-box \alg experiments. All ranged parameters sampled uniformly.}\label{tab:hyper} 
\begin{tabular}{@{}ll@{}}
\toprule
\textbf{Parameter}                                                                                         & \textbf{Value}          \\ \midrule
RGF smoothing parameter ($\beta$)                                                                             & 1                       \\
RGF step size ($\eta$)                                                                                        & 500                     \\
Boosting query budget ($b$)                                                                                  & 20k                     \\
\# Samples for RGF gradient sampling ($q$)                                                                   & 10                      \\ \midrule
Rotation range about the $y$ axis                                                                            & (-50$\degree${}, 50$\degree${}) \\
Base focal length ($f$)                                                                                      & 3 ft.                   \\
Crop sizes and offsets                                                                                     & (-3.125\%, 3.125\%)     \\
Lighting gamma value ($\gamma$)                                                                               & (1, 3.5) and (1, 1/3.5) \\
Gaussian kernels for blurring                                                                              & {[}1, 3, 5, 7{]}        \\ \midrule
\begin{tabular}[c]{@{}l@{}}Joint optimization mask size weight ($\lambda$)\end{tabular}                 & 5                       \\
\# of transforms ($n$)                                                                                       & 100                     \\
\begin{tabular}[c]{@{}l@{}}Transform-robustness high threshold\\  \hspace{1em}   for coarse red. ($tr_{hi}$)\end{tabular} & 85\%                    \\
\begin{tabular}[c]{@{}l@{}}Transform-robustness low threshold\\  \hspace{1em}   for fine red. ($tr_{lo}$)\end{tabular}    & 65\%                    \\
Max mask size option ($m_{max}$)                                                                                       & Not set          \\ \bottomrule
\end{tabular}
%\vspace{-0.1in}
\end{table}

                   \begin{table}
            \centering
            \caption{\label{tab:digital} Quantitative results of digital GTSRB attacks. Average and standard deviation of \eotmetric, \# of pixels out of $32 \times 32 = 1024$ in the mask (i.e., $\ell_0$ distance), number of queries reported for the given victim to all other targets.}
        
            \begin{tabular}{ >{\centering\arraybackslash}m{1.55cm}>{\centering\arraybackslash}m{1.6cm}>{\centering\arraybackslash}m{1.6cm}>{\centering\arraybackslash}m{1.6cm} } 
                 \toprule
                    \textbf{Victim} & \texttt{Stop} & \texttt{SL 30 km/hr}  & Avg. of all 43 victims\\\midrule
                    \textbf{TR (\%)} & 80.8$\pm$7.02 & 84.5$\pm$10.6 & 77.8$\pm$17.5\\
                    \textbf{Mask Size} & 116$\pm$38.0 & 139$\pm$71.7 & 170.3$\pm$122\\
                    \textbf{\# Queries} & 133k$\pm$8.92k & 134k$\pm$15.2k & 126k$\pm$18.9k\\\bottomrule

                \end{tabular}
        \vspace{-0.1in}
        \end{table}

        \newcommand{\tabwid}[0]{0.05\linewidth}
         \begin{table*}
            \centering
            \caption{ Sample of digital targeted attacks on GTSRB-Net. Ran with $tr_{lo} = 65\%$, $tr_{hi} = 85\%$, $n = 100$ transforms. For cells with same victim and target, we report the \% of transforms the original label is predicted. Masks size is reported in terms of \# of pixels in $32 \times 32$ (i.e., $\ell_0$ in $32 \times 32$ space). }\label{fig:dig_pics}
        
            \begin{tabular}{ c| cccccccccc } 
                 \toprule
                    & \multicolumn{10}{c}{{\large \textbf{Target}}}  \\
                    {\large \textbf{Victim}} &
                    \includegraphics[width=\tabwid]{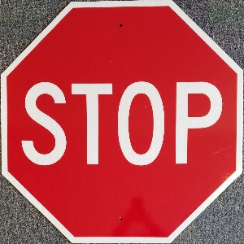}&
                    \includegraphics[width=\tabwid]{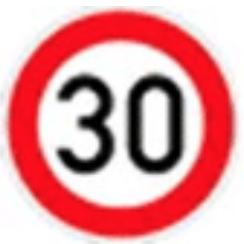} & \includegraphics[width=\tabwid]{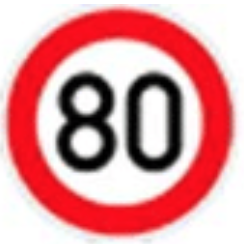} & \includegraphics[width=\tabwid]{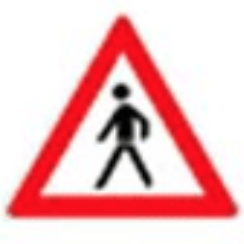}& \includegraphics[width=\tabwid]{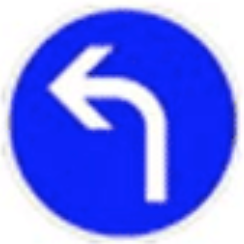}& \includegraphics[width=\tabwid]{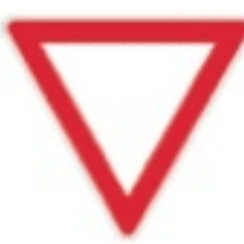}& \includegraphics[width=\tabwid]{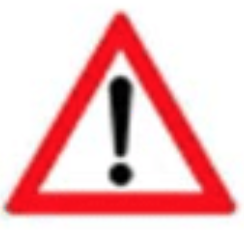}& \includegraphics[width=\tabwid]{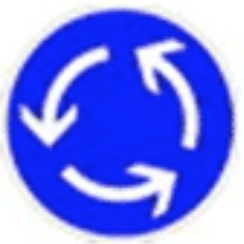}& \includegraphics[width=\tabwid]{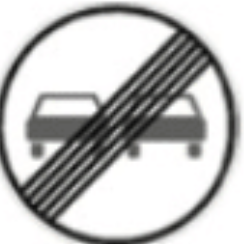}&  \includegraphics[width=\tabwid]{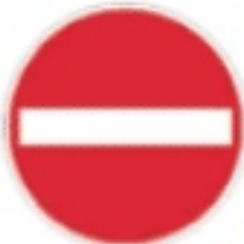}\\ \midrule
                    
                    \includegraphics[width=\tabwid]{figs/digital/14.png} &
                    \includegraphics[width=\tabwid]{figs/digital/14.png} &
                    \includegraphics[width=\tabwid]{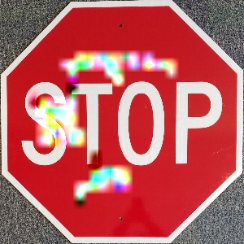} &
                    \includegraphics[width=\tabwid]{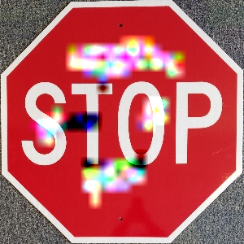} &
                    \includegraphics[width=\tabwid]{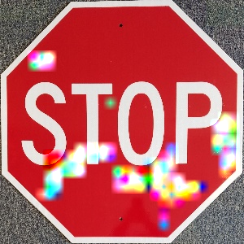} & \includegraphics[width=\tabwid]{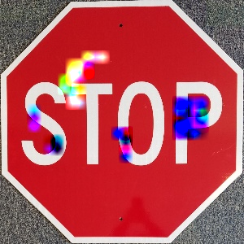} & \includegraphics[width=\tabwid]{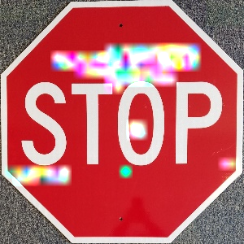} & \includegraphics[width=\tabwid]{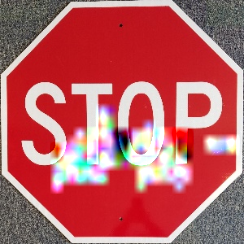} & \includegraphics[width=\tabwid]{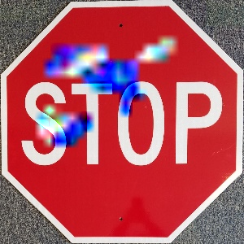} & \includegraphics[width=\tabwid]{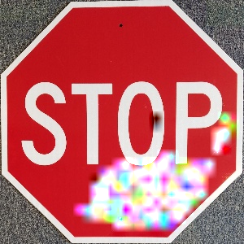} & \includegraphics[width=\tabwid]{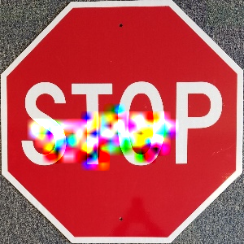}\\
                    {\em \EOTmetric} & 92\% & 86\% & 85\% & 79\% & 80\% & 82\% & 80\% & 90\% & 77\% & 71\%\\
                    {\em Mask Size} & 0 & 88 & 123 & 105 & 51 & 96 & 131 & 92 & 138 & 97 \\
                   % {\em Mask to Object Size Ratio} & 7.26\% & 14.96\% & 7.26\% & 12.53\%\\ \midrule
                   \includegraphics[width=\tabwid]{figs/digital/1.png} &
                   \includegraphics[width=\tabwid]{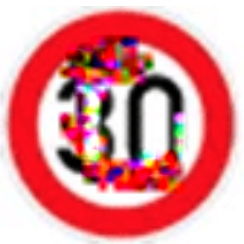} &
                    \includegraphics[width=\tabwid]{figs/digital/1.png} & \includegraphics[width=\tabwid]{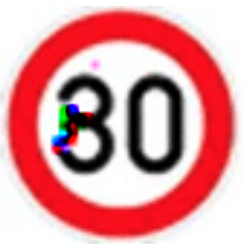} & \includegraphics[width=\tabwid]{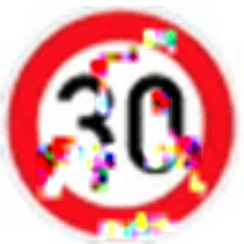} & \includegraphics[width=\tabwid]{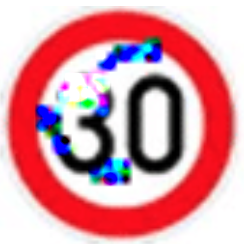} & \includegraphics[width=\tabwid]{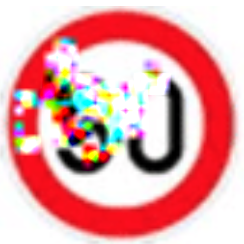} & \includegraphics[width=\tabwid]{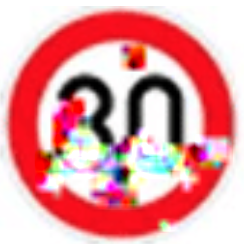} & \includegraphics[width=\tabwid]{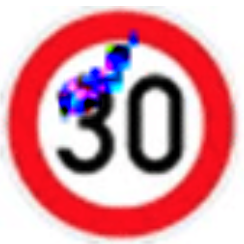} & \includegraphics[width=\tabwid]{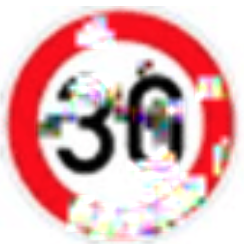} & \includegraphics[width=\tabwid]{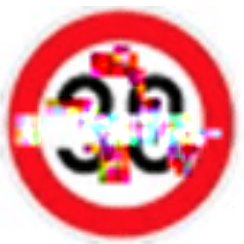} \\
                    {\em \EOTmetric} & 77\% & 100\% & 88\% & 94\% & 94\% & 89\% & 88\% & 99\% & 61\% & 93\%\\
                    {\em Mask Size} & 161 & 0 & 18 & 140 & 74 & 152 & 175 & 45 & 210 & 156 \\
                    %{\em Mask to Object Size Ratio} & 2.11\% & 21.08\% & 11.14\% & 0\%\\
                    
                    \includegraphics[width=\tabwid]{figs/digital/5.png} &
                   \includegraphics[width=\tabwid]{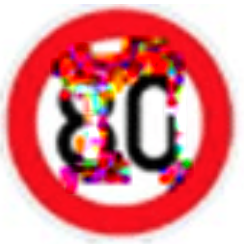} &
                    \includegraphics[width=\tabwid]{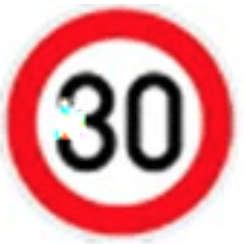} & \includegraphics[width=\tabwid]{figs/digital/5.png} & \includegraphics[width=\tabwid]{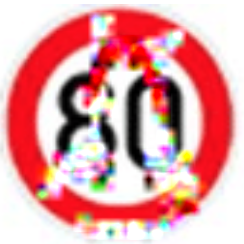} & \includegraphics[width=\tabwid]{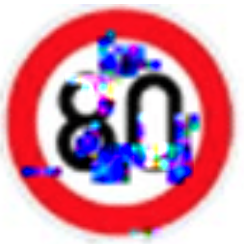} & \includegraphics[width=\tabwid]{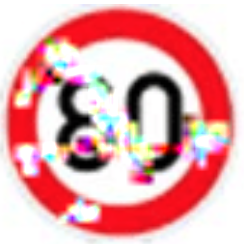} & \includegraphics[width=\tabwid]{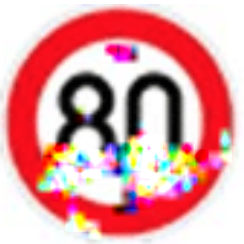} & \includegraphics[width=\tabwid]{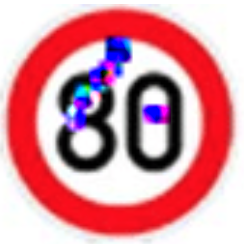} & \includegraphics[width=\tabwid]{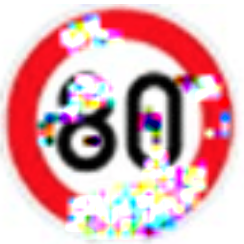} & \includegraphics[width=\tabwid]{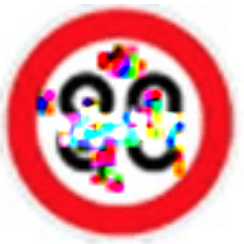} \\
                    {\em \EOTmetric} & 79\% & 97\% & 88\% & 87\% & 88\% & 91\% & 95\% & 97\% & 62\% & 65\%\\
                    {\em Mask Size} & 171 & 13 & 0 & 228 & 132 & 173 & 162 & 41 & 213 & 122 \\
                    
                    \includegraphics[width=\tabwid]{figs/digital/27.png} &
                   \includegraphics[width=\tabwid]{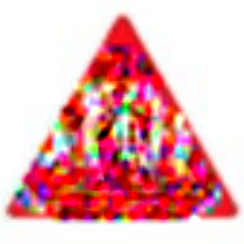} &
                    \includegraphics[width=\tabwid]{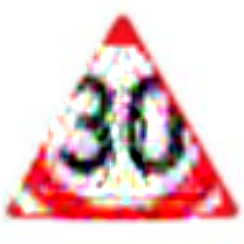} & \includegraphics[width=\tabwid]{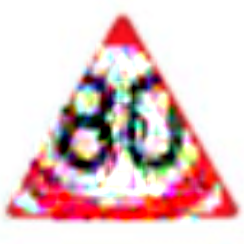} & \includegraphics[width=\tabwid]{figs/digital/27.png} & \includegraphics[width=\tabwid]{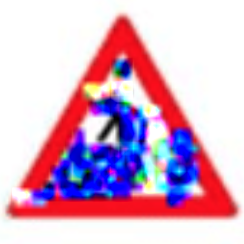} & \includegraphics[width=\tabwid]{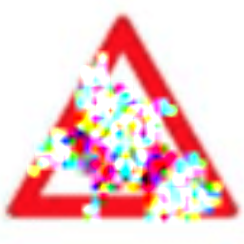} & \includegraphics[width=\tabwid]{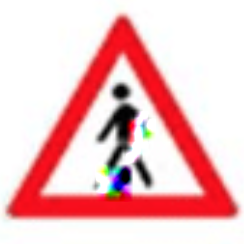} & \includegraphics[width=\tabwid]{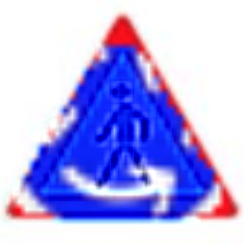} & \includegraphics[width=\tabwid]{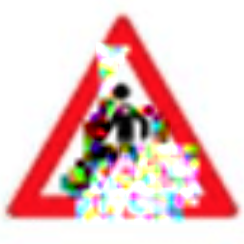} & \includegraphics[width=\tabwid]{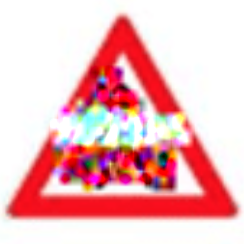} \\
                    {\em \EOTmetric} & 50\% & 73\% & 52\% & 95\% & 86\% & 79\% & 100\% & 0\% & 79\% & 88\%\\
                    {\em Mask Size} & 398 & 398 & 398 & 0 & 198 & 254 & 28 & 398 & 205 & 190 \\
                    
                    \includegraphics[width=\tabwid]{figs/digital/34.png} &
                   \includegraphics[width=\tabwid]{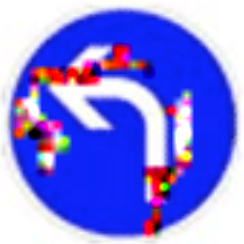} &
                    \includegraphics[width=\tabwid]{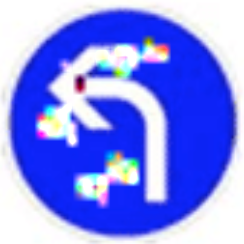} & \includegraphics[width=\tabwid]{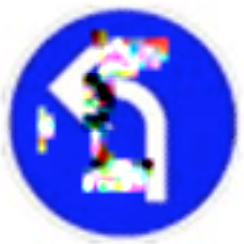} & \includegraphics[width=\tabwid]{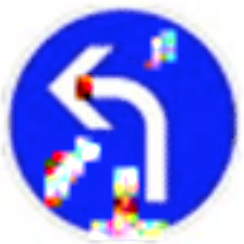} & \includegraphics[width=\tabwid]{figs/digital/34.png} & \includegraphics[width=\tabwid]{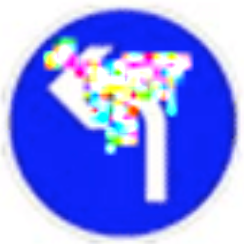} & \includegraphics[width=\tabwid]{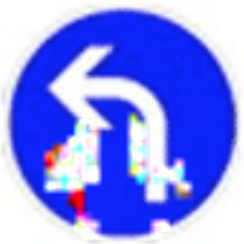} & \includegraphics[width=\tabwid]{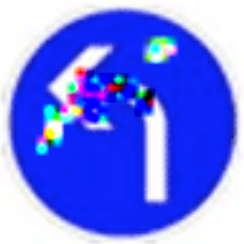} & \includegraphics[width=\tabwid]{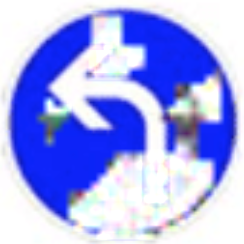} & \includegraphics[width=\tabwid]{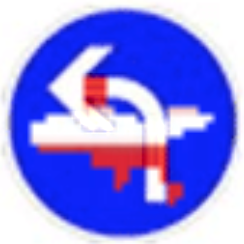} \\
                    {\em \EOTmetric} & 73\% & 85\% & 71\% & 85\% & 100\% & 81\% & 71\% & 80\% & 71\% & 71\%\\
                    {\em Mask Size} &111 & 72 & 120 & 92 & 0 & 122 & 93 & 74 & 263 & 148 \\
                    
                    \includegraphics[width=\tabwid]{figs/digital/13.png} &
                   \includegraphics[width=\tabwid]{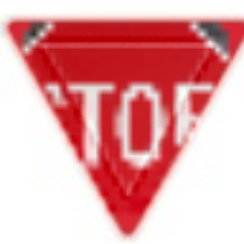} &
                    \includegraphics[width=\tabwid]{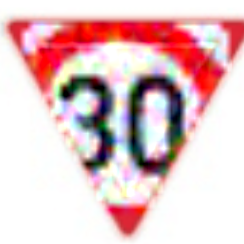} & \includegraphics[width=\tabwid]{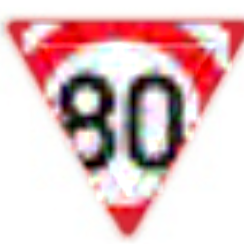} & \includegraphics[width=\tabwid]{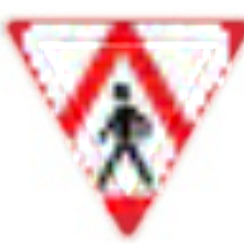} & \includegraphics[width=\tabwid]{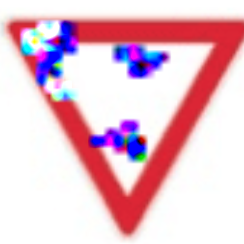} & \includegraphics[width=\tabwid]{figs/digital/13.png} & \includegraphics[width=\tabwid]{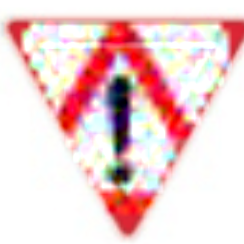} & \includegraphics[width=\tabwid]{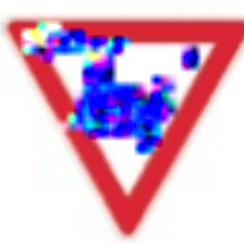} & \includegraphics[width=\tabwid]{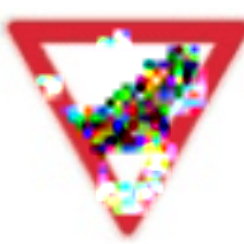} & \includegraphics[width=\tabwid]{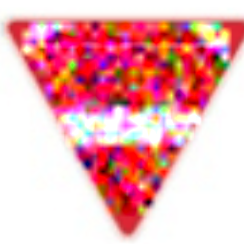} \\
                    {\em Tranform robustness} & 0\% & 58\% & 7\% & 4\% & 92\% & 100\% & 46\% & 87\% & 83\% & 57\%\\
                    {\em Mask Size} &436 & 436 & 436 & 436 & 81 & 0 & 436 & 157 & 198 & 436 \\
                    
                    \includegraphics[width=\tabwid]{figs/digital/18.png} &
                   \includegraphics[width=\tabwid]{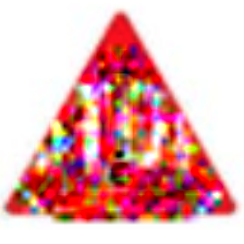} &
                    \includegraphics[width=\tabwid]{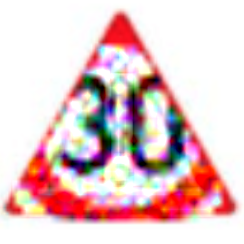} & \includegraphics[width=\tabwid]{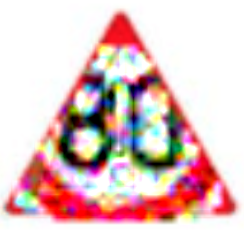} & \includegraphics[width=\tabwid]{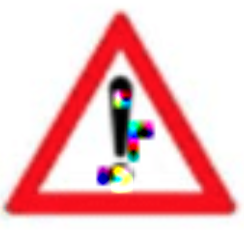} & \includegraphics[width=\tabwid]{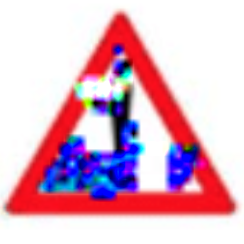} & \includegraphics[width=\tabwid]{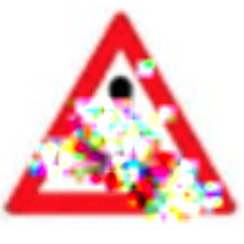} & \includegraphics[width=\tabwid]{figs/digital/18.png} & \includegraphics[width=\tabwid]{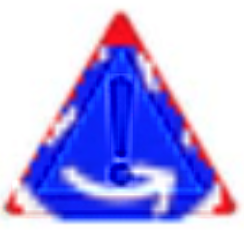} & \includegraphics[width=\tabwid]{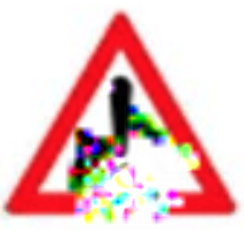} & \includegraphics[width=\tabwid]{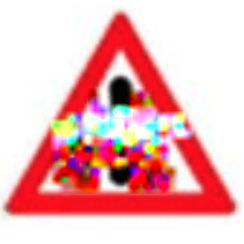} \\
                    {\em Tranform robustness} & 56\% & 84\% & 70\% & 90\% & 80\% & 69\% & 100\% & 0\% & 78\% & 77\%\\
                    {\em Mask Size} &398 & 398 & 398 & 23 & 133 & 225 & 0 & 398 & 133 & 161 \\
                    
                    \includegraphics[width=\tabwid]{figs/digital/40.png} &
                   \includegraphics[width=\tabwid]{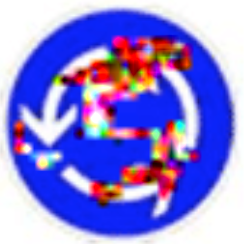} &
                    \includegraphics[width=\tabwid]{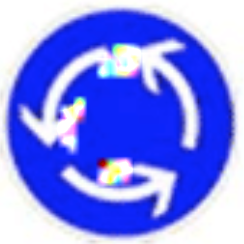} & \includegraphics[width=\tabwid]{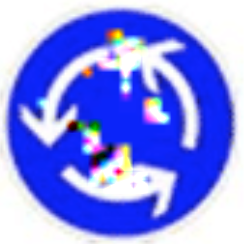} & \includegraphics[width=\tabwid]{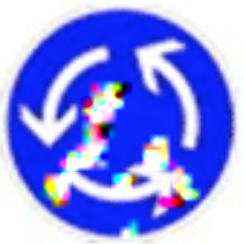} & \includegraphics[width=\tabwid]{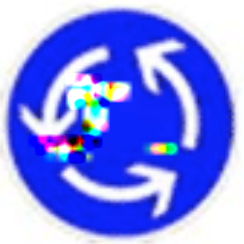} & \includegraphics[width=\tabwid]{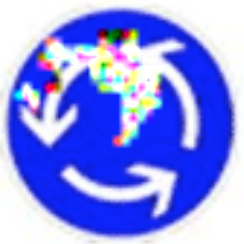} & \includegraphics[width=\tabwid]{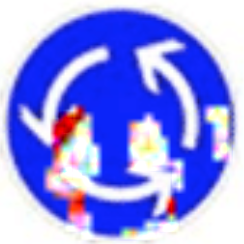} & \includegraphics[width=\tabwid]{figs/digital/40.png} & \includegraphics[width=\tabwid]{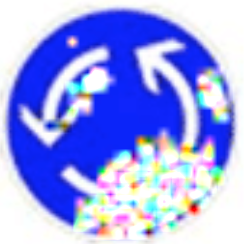} & \includegraphics[width=\tabwid]{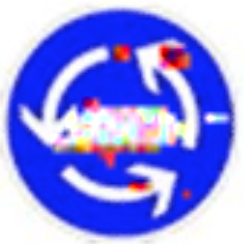} \\
                    {\em \EOTmetric} & 74\% & 91\% & 73\% & 86\% & 77\% & 78\% & 76\% & 99\% & 72\% & 69\%\\
                    {\em Mask Size} & 172 & 43 & 75 & 92 & 62 & 106 & 133 & 0 & 218 & 114 \\
                    
                    \includegraphics[width=\tabwid]{figs/digital/41.png} &
                   \includegraphics[width=\tabwid]{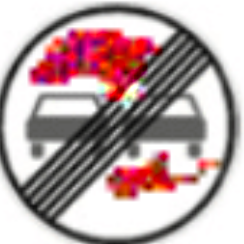} &
                    \includegraphics[width=\tabwid]{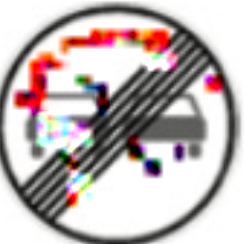} & \includegraphics[width=\tabwid]{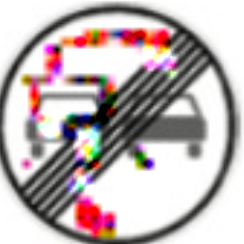} & \includegraphics[width=\tabwid]{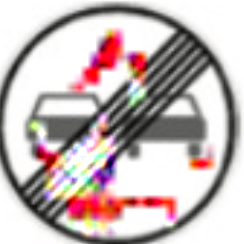} & \includegraphics[width=\tabwid]{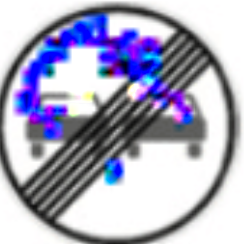} & \includegraphics[width=\tabwid]{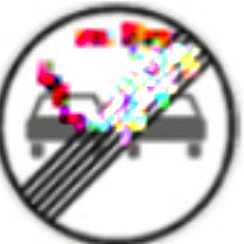} & \includegraphics[width=\tabwid]{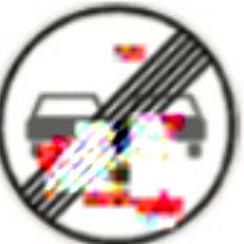} & \includegraphics[width=\tabwid]{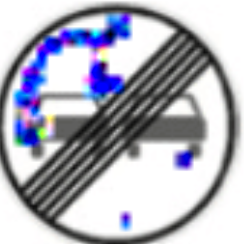} & \includegraphics[width=\tabwid]{figs/digital/41.png} & \includegraphics[width=\tabwid]{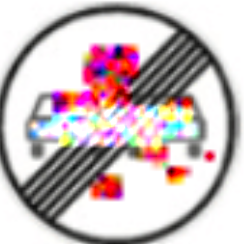} \\
                    {\em \EOTmetric} & 72\% & 83\% & 82\% & 74\% & 91\% & 92\% & 78\% & 90\% & 73\% & 79\%\\
                    {\em Mask Size} & 142 & 123 & 141 & 159 & 137 & 137 & 175 & 90 & 0 & 182 \\
                    
                    \includegraphics[width=\tabwid]{figs/digital/17.png} &
                   \includegraphics[width=\tabwid]{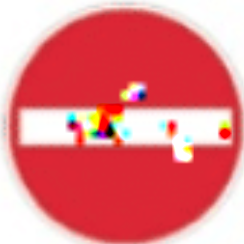} &
                    \includegraphics[width=\tabwid]{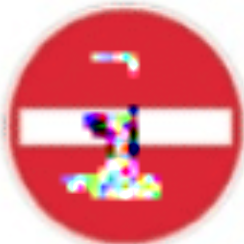} & \includegraphics[width=\tabwid]{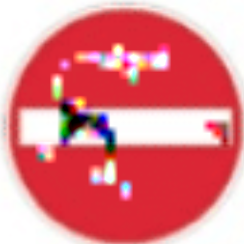} & \includegraphics[width=\tabwid]{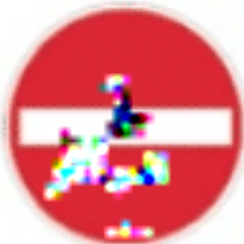} & \includegraphics[width=\tabwid]{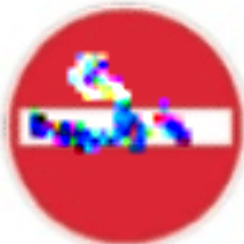} & \includegraphics[width=\tabwid]{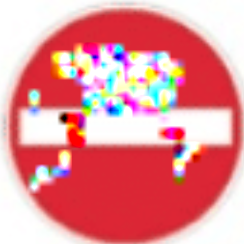} & \includegraphics[width=\tabwid]{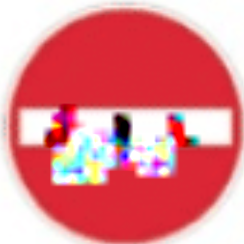} & \includegraphics[width=\tabwid]{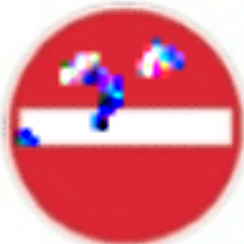} & \includegraphics[width=\tabwid]{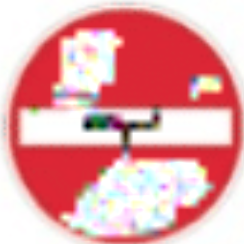} & \includegraphics[width=\tabwid]{figs/digital/17.png} \\
                    {\em \EOTmetric} & 79\% & 89\% & 89\% & 81\% & 82\% & 67\% & 82\% & 91\% & 73\% & 99\%\\
                    {\em Mask Size} & 42 & 85 & 79 & 106 & 106 & 153 & 117 & 62 & 242 & 0 \\
                 \bottomrule
                \end{tabular}
        \vspace{-0.15in}
        \end{table*}

As in our white-box experiments in Section~\ref{sec:white-box}, we use Internet images outside of the dataset plus our own \texttt{Stop sign} picture since we had a physical sign available for initialization to demonstrate that \alg does not rely on having training set images to initialize from. In our experiments we assume that object boundaries are available but note they could be obtained automatically through an object segmentation network~\cite{objectsegmentation}.

\textbf{GTSRB Attack Details.}
    We set the size of our input images to be 244$\times$244. During the attack, we generate 32$\times$32 perturbations and then upsample the perturbations to the resolution of the input image when they are added. Reducing the dimensionality of the perturbation space makes RGF more efficient~\cite{cheng2018query} and can help encourage blockier perturbations that can be sensed by a camera in the physical-world.

        For field testing, we print stickers and place them on a $30" \times 30"$ \texttt{Stop sign} and place it at stationary positions to test how robust our attacks are to different viewing conditions. We take five pictures of the perturbed \texttt{Stop sign} at 14 different locations for a total of 70 pictures per set. To test lighting conditions, we take one set of images in outdoor light and two sets indoors, one with indoor lights on and one without. To compare against baseline \texttt{Stop sign} accuracy, we also take five pictures of a clean \texttt{Stop sign} at each of the same 14 locations. The 14 locations where chosen based on RP${}_2$ evaluation~\cite{roadsigns17}. 
        
        To gather crops, we use the original YOLOv3~\cite{redmon2018yolov3} object detector network trained on MS COCO~\cite{lin2014microsoft} to predict bounding boxes for the \texttt{Stop sign}. We take the output bounding boxes, crop the sign, resize to $32 \times 32$, and feed through our network for classification. Further hyperparameters are specified in Table~\ref{tab:hyper}.

\subsection{Experimental Results}\label{sec:exp}
We first report digital results for GTSRB and CIFAR-10. We then report physical transform-robustness field tests results on GTSRB\footnote{CIFAR-10 objects such as airplanes or birds would have been difficult for applying physical perturbations.}. Finally, we report physical drive-by results on GTSRB.
%\textbf{Baseline Comparisons.}
%We compare our results to digital OPT-attack and OPT+EoT baselines. For the OPT+EoT baseline, we take digital OPT-attack and average the gradient direction over 100 transformations and otherwise leave the attack the same. The results for 3 attacks, \texttt{Stop sign} to \texttt{Speed Limit 30 km/hr}, \texttt{Stop sign} to \texttt{Pedestrians}, and \texttt{Stop sign} to \texttt{Turn Left Ahead}, are shown in Table~\ref{tab:digital_baselines}.

%\alg generally uses a comparable amount of queries compared to either baseline, despite using most of its queries on its extra mask generation step. The \alg results in Table~\ref{tab:digital_baselines} only used 20k queries in boosting. Compared to the digital OPT-attack baseline, \alg also solves a much harder problem in yielding physically survivable results. We also point out that the OPT+EoT baseline yields much worse artifacts across the entire image of the target image compared to \alg without using significantly fewer queries in most cases. Thus, \alg is much more suited to this problem than the naive OPT + EoT baseline.

\textbf{Digital Transform-Robustness Results (GTSRB).}
We report results for all 1,806 possible GTSRB victim target pairs in Table~\ref{tab:digital}. On average, we observe masks with an $\ell_0$ distance of 170.3 (16.6\% w.r.t. the $32 \times 32$ image area) and 77.8\% \eotmetric.

%%%% OLD digital pictures table
            \begin{table*}
            \centering
            \caption{\label{tab:physical_table_gtsrb} GTSRB field test results. Physical robustness results are calculated over 5 pictures each at the following spots: 5 ft $\times$ \{0\degree, 15\degree, 30\degree, 45\degree\}, 10 ft $\times$ \{0\degree, 15\degree, 30\degree\}, 15 ft $\times$ \{0\degree, 15\degree\}, 20 ft $\times$ \{0\degree, 15\degree\}, 25 ft, 30 ft, 40 ft. Each example was tested 3 times: outdoors, indoors with indoor lights turned off, and indoors with indoor lights turned on. }
        
            \begin{tabular}{ >{\centering\arraybackslash}m{2cm}>{\centering\arraybackslash}m{1.8cm}>{\centering\arraybackslash}m{2cm}>{\centering\arraybackslash}m{2cm}>{\centering\arraybackslash}m{1.5cm}>{\centering\arraybackslash}m{1.5cm}>{\centering\arraybackslash}m{1.5cm}>{\centering\arraybackslash}m{1.5cm} } 
                 \toprule
                    \textbf{Victim} & \textbf{Target}& \textbf{Digital GRAPHITE attack}& \textbf{Physical GRAPHITE attack (outdoors)}&
                    \textbf{Dig. TR (100 xforms)}&\textbf{Phys. TR (Indoors, lights off)}& \textbf{Phys. TR (Indoors, lights on)}& \textbf{Phys. TR (Outdoors)}\\ \midrule
                   \includegraphics[width=0.85\linewidth]{figs/digital/14.png} & \includegraphics[width=0.85\linewidth]{figs/digital/1.png} &\includegraphics[width=0.85\linewidth]{figs/digital/14_1.png}& \includegraphics[width=0.85\linewidth]{figs/14_1_outdoor.jpg} & 86\% &92.9\% & 94.3\% & 100\%\\
                   \includegraphics[width=0.85\linewidth]{figs/digital/14.png} & \includegraphics[width=0.85\linewidth]{figs/digital/27.png}&\includegraphics[width=0.85\linewidth]{figs/digital/14_27.png}& \includegraphics[width=0.85\linewidth]{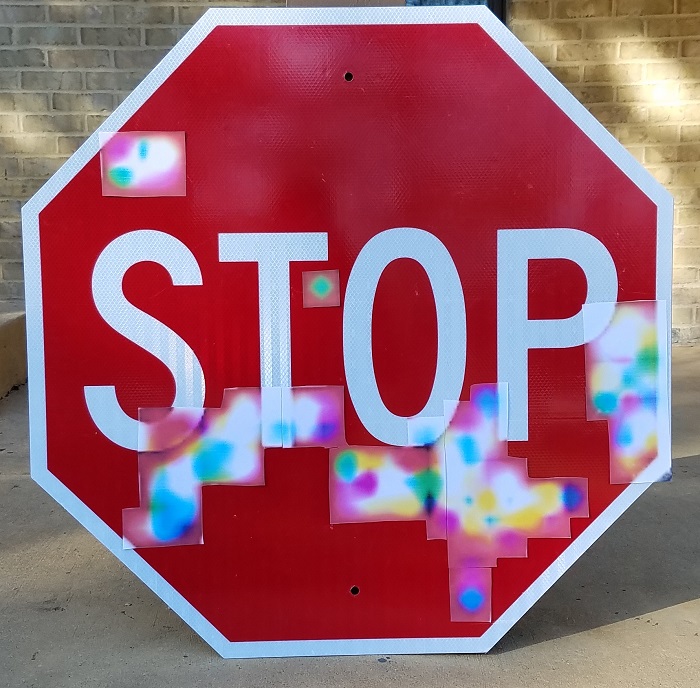} & 79\% &97.1\% & 85.7\% & 100\%\\

                 \bottomrule
                \end{tabular}
                \vspace{-0.1in}
            \end{table*}

In Table~\ref{fig:dig_pics} we provide results from attacks between all victim target pairs of a subset of 10 varied GTSRB signs. We include the final output image, the \eotmetric, and the mask size. \alg generally finds small, highly transform-robust perturbations on a variety of attacks. Examples that tend to not perform well are attacks on a triangle sign victim, as overcoming the difference in shape is difficult. Furthermore, it is not large enough to fit the entire target object on the victim. This means that in reduction the starting \eotmetric may be low to begin with, making it hard to prune patches from the mask. Resizing the target sign to fit within the victim sign may alleviate this. Generally, the attack quality varies depending on the victim target pair, which is consistent with prior work that finds that the distortion varies drastically depending on the attack pair~\cite{papernot2016limitations}. 

\textbf{Digital Transform-Robustness Results (CIFAR-10).} We test on 500 random victim-target pairs from CIFAR-10~\cite{krizhevsky2009learning} on Wide ResNet 34-10~\cite{zagoruyko2016wide}, commonly used to evaluate CIFAR-10 attacks and defenses~\cite{madry2017towards,zhang2019theoretically,zheng2020efficient}. We observe that even natural images are not robust to the same set of transformations used in GTSRB, so we reduce the perspective transform to [-30, 30] at a constant focal length, reduce $\gamma$ to (1, 2) and (1, 1/2), and remove the blur. We achieve an average transform-robustness of 87.43\% and mask size of 125.5 with 141.9k queries. We include example attacks in Appendix~\ref{ap:cifar}.

%%%%%%%%%%%%%%%%%%%%% Old physical table location

 \iffalse
\begin{figure}[t]
    \centering
    \includegraphics[width=0.9\linewidth]{figs/survivaltime.png}
    \caption{The above plot shows the three phases of the algorithm. The plot starts near time 286 because it takes approximately 286 seconds to build up a heatmap. The binary search stage is very quick and reduces the number of pixels in the mask from 702 to 215 with a \eotmetric of 90\%. Over the next 113 seconds, the reduce stage further reduces the number of bits in the mask to 70 bits and approximate \eotmetric to 76\%. The boost stage starts at approximately 475 seconds. It boosts the \eotmetric to 93\% while keeping the same number of bits.
    further reduces the number of bits to 70 with a \eotmetric of 70\%.}
    \label{fig:survivabilitytime}
\end{figure}
\fi

%-------------------------------------------------------------------------

\textbf{Physical Transform-Robustness Field Tests (GTSRB).}
We also conduct field experiments to confirm that our results carry over to the physical world. This consists of printing stickers and placing them on real-world traffic signs and then evaluating images at differing conditions. We evaluate \alg on targeted \texttt{Stop sign} attacks to \texttt{Speed Limit 30 km/hr} and \texttt{Pedestrians} at different viewing angles and lighting.

   Table~\ref{tab:physical_table_gtsrb} shows the results of the field tests for GTSRB. On average, they took 131k queries. The \texttt{Speed Limit 30 km/hr} attack was successful over all three lighting conditions with at least 92.9\% physical robustness. The \texttt{Pedestrians} attack also did well, with at least 85.7\% success rate over each lighting condition. Overall, these results suggest that \eotmetric translates reasonably well into the physical world and improvements in transformations could further improve the translation. Baseline \texttt{Stop sign} tests were at least 95.7\% in each lighting condition. % We include extended field test results that characterize attack success under additional imaging conditions in Section~\ref{sec:supp_drive} of the supplementary material.

      %  Overall, 92.86\% of the 140 total images (one set of five at each location mid-day and another set in the evening) were classified as speed limit 30 km/hr. There was not too much variation across either lighting changes or geometric changes, with all spots yielding a successful attack image at least 80\% of the time and with little difference between mid-day and evening photos (94.29\% vs 91.43\%). These results suggest that our perturbations are robust to environmental changes captured by viewing perspectives and lighting. The average confidence is above 85\% as well. We also confirmed that the speed limit 30 km/hr classifications were due to our stickers with a separate baseline test suite on a clean Stop sign. In the baseline, all 100\% of the clean Stop sign images were correctly classified. 

\textbf{Physical Drive-by Tests (GTSRB).}
    In this section, we provide extended field test results for GTSRB under additional imaging conditions, in the form of drive-by tests. %To conduct this test, we had one of our authors use a real car to drive by the sign, as if the vehicle were approaching the sign and needed to classify it correctly. We used the experimental method from prior work~\cite{roadsigns17}. 
    In particular, we record videos while driving towards the sign in a private lot, simulating a realistic driving environment in an allowable fashion\footnote{We follow local laws and regulations for safety in this test.}. We test each of the two attacks from Table~\ref{tab:physical_table_gtsrb}. As in prior work by Eykholt et al. ~\cite{roadsigns17}, we analyze every 10th frame from the video. We use the same YOLOv3~\cite{redmon2018yolov3} detector as the other GTSRB results to crop our frames. Some frames, particularly ones at the beginning of the video when we were farthest away could not be cropped with YOLOv3 and were thus cropped manually in a similar fashion.

                 \begin{table}
            \centering
            \caption{\label{tab:drive-by-results} Extended GTSRB field test results, including the measured physical robustness of the video frames. Every 10th frame of the videos were analyzed. }
        
            \begin{tabular}{ >{\centering\arraybackslash}m{2.5cm}>{\centering\arraybackslash}m{2.cm}>{\centering\arraybackslash}m{2.25cm} } 
                 \toprule
                    \textbf{Target}& \textbf{\# Frames Analyzed}& \textbf{Attack Success Rate (Transform-robustness)} \\ \midrule
                  \texttt{Speed Limit 30 km/hr} & 40 & 97.5\% \\ 
                 \texttt{Pedestrians} & 39 & 82.1\% \\
                 \bottomrule
                \end{tabular}
                \vspace{-0.1in}
            \end{table}
            
    We include results for these two stop sign attacks in Table~\ref{tab:drive-by-results}. The \texttt{Speed Limit 30 km/hr} attack was the most successful, with 97.5\% physical robustness, which is in line with the field test results in Table~\ref{tab:physical_table_gtsrb} and above the digital \eotmetric. The \texttt{Pedestrians} attack was also successful with an 82.1\% physical robustness rate, right around its digital \eotmetric of 79\%.
    
    Overall, these results confirm the field test results reported in Table~\ref{tab:physical_table_gtsrb} in an even more realistic driving setting. Example images are included in Appendix~\ref{ap:drive-by}.

\input{tex/variations}
\if\includedriveby

\subsection{Drive-by}
  \subsubsection{Drive-by Test setup}
        Boosting can be used alone on a pre-selected mask. We did a drive-by test using a pre-selected mask and then boosting it as in this paper.  To evaluate our attack in a driving scenario, similar to Eykholt \etal~\cite{roadsigns17} we record videos of us driving up to the stop sign with a mounted smartphone camera in an empty parking lot. We found  98\% of every 10th frames analyzed to be classified as 30 km/hr. In contrast, without perturbation applied, the classifier correctly classified the stop sign in all the frames. The adversarial patch applied to a physical stop sign in our tests is shown in Figure~\ref{fig:drive-by}. This drive-by test was done pre-Covid-19 and before the mask generation algorithm was in place. Later experiments with  GRAPHITE's patches automatic mask generation were done in one of the author's backyard (Tables~\ref{tab:lab-adv} and \ref{tab:lab-qualitative}).

   %\subsubsection{Drive-by Test Results}
     %   We analyze every 10th frame by cropping out the sign and passing it to our classifier. On a Samsung phone video 39 frames were classified as speed limit 30 km/hr with an average confidence of 0.87. On an iPhone video, 29 out of the 30 frames were classified as speed limit 30 km/hr, with an average confidence of 0.93 on those 29 frames. In a clean stop sign video, all analyzed frames were labeled with stop sign. Example pictures can be shown in Fig.~\ref{fig:drive-by}.
       
\begin{figure}
    \centering
    \includegraphics[width=0.9\linewidth]{figs/drive-by-examples.PNG}
    \caption{Stop sign to speed limit 30 km/hr drive-by images}
    \label{fig:drive-by}
\end{figure}
\fi

%% file: tex/variations.tex
\subsection{Variations and Tuning of \alg}\label{sec:extra_results}

We now explore \alg's ability to trade-off \eotmetric, mask size, and query count as well as an examination of heatmap estimation strategies. These tradeoffs allow an attacker to design their attack given the relative importance of certain attack constraints, such as a need to keep the perceptibility of such attacks low or a need to keep the query cost under a certain budget. We also examine \alg's output without the restriction that the noise be contained within the victim object and a strategy with multiple mask generation / boosting rounds.

\subsubsection{Reducing the Number of Queries}
To further reduce the number of queries, we consider two adjustments: 1) reducing the number of transforms and 2) replacing the target-based heatmap strategy with a random heatmap strategy that simply orders patches randomly. We also test a ``minimum query" setting, which uses both adjustments to the extreme: both the random heatmap and only one transform.

\textbf{Tuning the Number of Transformations.}
We first begin by decreasing $n$ (the number of transformations) that we use in mask generation and boosting. This makes every \eotmetric measurement run faster. We test with $n=10, 20, 30, 40, 50, 60, 70, 80, 90, 100$ transforms. Regardless of the value of $n$, we measure the final transform-robustness of each setting with 100 transformations to ensure the numbers are comparative. We test the 90 non-identity victim-target pairs from Table~\ref{fig:dig_pics}. The results of these tests are shown in Fig.~\ref{fig:tune_xform}. Since the number of queries scales monotonically with the number of transforms, we plot the 10 points using the number of queries on the $x$ axis, and with the \eotmetric and mask size on the two $y$ axes. 

\input{tex/variations_figure}

% \begin{figure}
%     \centering
%     \includegraphics[width=0.5\textwidth]{figs/tune_xform.png}

% \caption{\EOTmetric and mask size vs. number of queries as the number of transformations is varied from [10, 20, 30, 40, 50, 60, 70, 80, 90, 100], on 90 victim-target pairs. Generally, as the number of queries and transformations are increased, the mask size decreases and the \eotmetric increases.}\label{fig:tune_xform}
% \end{figure}

As more transforms are used in mask generation and boosting, we find that the quality of the generated attack increases, i.e., the size of the mask decreases and \eotmetric increases. Intuitively, this makes sense - as the number of transformations are increased, the objective becomes smoother, and the estimates become more accurate. This also explains why the data at points with fewer transformations are noisy within the relative scale of the plot, as the sampled transformations may represent the true distribution poorly. This performance benefit comes at the trade-off of increased query count and thus, if given a specific query budget, an attacker could tune this parameter based on their query restrictions.

\textbf{Random Heatmap.}
For this variation, we consider replacing the target-based heatmap estimation process with a random heatmap strategy that simply orders the patches randomly, processing the patches in fine-grained reduction in an arbitrary order (note that we remove coarse-grained reduction, as its use of binary search makes it no longer well-defined). This significantly reduces the queries by saving an iteration over patch transform-robustness estimations. While we lose the ability to binary search, we save several thousand queries with this change. The objective function of~\eqref{eq:mask} rejects the removal of extremely bad choice of patches but can suffer from poor removal choices early on that negatively influence  the final result.

       \begin{table}
            \centering
            \caption{\label{tab:random_results} Comparison of the original target-based heatmap vs. a random heatmap strategy. The number of queries can be drastically decreased with, on average, some increase in mask size and some decrease in \eotmetric. Average and standard deviation are presented.}
        
            \begin{tabular}{ >{\centering\arraybackslash}m{1.6cm}>{\centering\arraybackslash}m{1.6cm}>{\centering\arraybackslash}m{1.6cm}>{\centering\arraybackslash}m{1.6cm} } 
                 \toprule
                    \textbf{Strategy} & Orig. Strategy with Target Heatmap & Random Heatmap & Min Query Setting\\\midrule
                    \textbf{TR (\%)} & 77.8$\pm$17.5 & 76.4$\pm$17.1 & 19.9$\pm$21.2 \\
                    \textbf{Mask Size} & 170.3$\pm$122 & 175.9$\pm$121 & 112.993$\pm$69.7\\
                    \textbf{\# Queries} & 126k$\pm$18.9k & 62.1k$\pm$12.5 & 572$\pm$105\\\bottomrule

                \end{tabular}
                \vspace{-0.1in}
        \end{table}

The results of this experiment are shown in Table~\ref{tab:random_results}, run over all 1,806 victim-target pairs. The random heatmap strategy performs quite similarly to the original target-based heatmap strategy with only about half of the queries, with the average number of queries dropping from 126k to 62.1k. The loss in attack quality is a drop from 77.8\% to 76.4\% \eotmetric and an increase from 170.3 to 175.9 pixels in terms of mask size. So, random heatmap is a viable strategy when query count is a concern. On average, however, the target-based heatmap still outperforms in terms of \eotmetric and mask generation.

\textbf{Min Queries Setting.}
We now combine both previous experiments by running \alg with the random heatmap and with only one transformation, to see if we can get a result that could possibly cause an accident in the right conditions with the minimum possible queries.

The results are shown in Table~\ref{tab:random_results}, run over all victim-target pairs. We find that, on average, \alg can find attacks with 19.9\% \eotmetric using only 572 queries. This includes 1,665 / 1,806 victim-target pairs with greater than 0\% \eotmetric. The mask size is also small, with an average of 112.993 pixels (11.0\% of the image), with fewer transformations for mask generation to have to optimize for. On just the 1665 non-zero \eotmetric examples, the average mask size was 107.897 pixels (10.5\% of the image) and the average query count was 566.253 queries. This suggests that risky examples could be generated in as few as a couple hundred queries but acquiring more robustness to a variety of transformations requires more queries.

\subsubsection{Reducing the Mask Size}
To reduce the mask size, we enable the $m_{max}$ option and vary it to the desired maximum mask size. This option can help achieve a desired mask size limit or make a less perceptible attack. We also change $tr_{lo} = 0$ to ensure that the maximum mask size can be reached, even if it means choosing a mask with a \eotmetric of 0.

% \begin{figure}
%     \centering
%     \includegraphics[width=0.5\textwidth]{figs/tune_mask.png}

% \caption{\EOTmetric and query count vs. mask size as $m_{min}$ is varied from [20, 41, 61, 82, 102, 123, 143, 164, 184, 205]. As $m_{min}$ increases, the \eotmetric increases and the query count decreases.}.\label{fig:tune_mask}
% \end{figure}

We test with $m_{max}$ values of 20, 41, 61, 82, 102, 123, 143, 164, 184, and 205, which corresponds to 2\%, 4\%, ..., 20\% of the image area. Since mask size increases monotonically as $m_{max}$ increases, we plot the actual average mask size on the $x$ axis. The results of these tests are shown in Fig.~\ref{fig:tune_mask}. The results show an increase in \eotmetric as the maximum mask size increases (and thus  the area to perturb increases) and a decrease in query count, as the algorithm can get below the maximum size quicker and exit earlier.

\subsubsection{Increasing \eotmetric}
To adjust \alg based on a desired transform-robustness level, we can tune the value of $tr_{lo}$. By increasing $tr_{lo}$, we can achieve higher levels of \eotmetric but potentially at the cost of larger mask size. This is because $tr_{lo}$ acts like a floor of acceptable \eotmetric while performing mask generation and guarantees a higher transform-robustness point on boosting. So, on average, a higher $tr_{lo}$ should result in higher \eotmetric.

% \begin{figure}
%     \centering
%     \includegraphics[width=0.5\textwidth]{figs/tune_slo.png}

% \caption{Mask size and query count vs. transform-robustness as $s_{lo}$ is varied from [0.1, 0.2, 0.3, 0.4, 0.5, 0.6, 0.7, 0.8, 0.9]. Generally, as $s_{lo}$ increases, the mask size increases as we reject more mask removals and the query count increases some as fewer overlapping patches are removed early.}.\label{fig:tune_slo}
% \end{figure}

We test with $tr_{lo}$ values of 0.1, 0.2, 0.3, 0.4, 0.5, 0.6, 0.7, 0.8, and 0.9. We fix $tr_{hi}$ to 1.0 for all values of $tr_{lo}$ so that it can remain constant across the tests. The average observed \eotmetric increased monotonically with $tr_{lo}$, so we plot the actual \eotmetric on the $x$ axis. The results of these tests are shown in Fig.~\ref{fig:tune_slo}. The results show an increase in both mask size and number of queries as \eotmetric increases, as expected. As such, this parameter can help raise transform-robustness if the attack is not as successful as need be in the physical-world. Conversely, lowering $tr_{lo}$ may help find smaller, less perceptible attacks. %Mask size is directly impacted with fewer patches being able to be removed, and query count is slightly impacted as fewer overlapping patches are removed early, forcing increased transform-robustness patch estimates.

\subsubsection{Removing the Victim Object Boundary Restrictions}
To examine \alg's performance in a setting closer to standard patch attacks, we remove the restriction that the perturbation be contained to the victim object.  

\begin{table}
            \centering
            \caption{\label{tab:full_hull_results} Comparison of the original victim-object constrained attack and an attack where the perturbation can be placed anywhere in the image. Average and standard deviation are presented.}
        
            \begin{tabular}{ >{\centering\arraybackslash}m{2.2cm}>{\centering\arraybackslash}m{2.2cm}>{\centering\arraybackslash}m{2.2cm} } 
                 \toprule
                    \textbf{Strategy} & Orig. Strategy, Restricted to Victim Object & Area Unrestricted \\\midrule
                    \textbf{TR. (\%)} & 77.8$\pm$17.5 & 86.7$\pm$8.81 \\
                    \textbf{Mask Size} & 170.3$\pm$122 & 147.1$\pm$77.8\\
                    \textbf{\# Queries} & 126k$\pm$18.9k & 149k$\pm$20.6\\\bottomrule

                \end{tabular}
                \vspace{-0.1in}
        \end{table}

The results show that \alg can perform even better if it is not concerned with fitting it in within the boundary of the victim object. The average \eotmetric increases from 77.8\% to 86.7\% and the mask size drops from 170.3 pixels ($\approx 16.6\%$ of the image) to 147.1 pixels ($\approx 14.4\%$ of the image). With more valid patches to evaluate, the query count raises slightly from 126k to 149k. Thus, if the threat model does not need to remain strictly in victim object, smaller attacks can be found with higher \eotmetric.

\subsubsection{Alternating Repeatedly between Mask Generation and Boosting}\label{subsec:joint}
We now apply multiple rounds of mask generation and boosting as an approximation of a joint optimization approach, testing the 90 non-identity victim-target pairs from Table~\ref{fig:dig_pics} on two rounds. We find that the average mask size improves from 172.0 to 138.9 pixels (a 19.3\% improvement) but transform-robustness decreases from 75.3\% to 74.2\% and query count increases from 129.5k to 213.3k queries (a 64.7\% increase). While this strategy can yield smaller masks, the query count increases faster than the performance improvements.

%% file: tex/variations_figure.tex
 \begin{figure*}[t]
     \centering
       \subfloat[Effects of tuning the number of transforms, on 90 victim-target pairs. Generally, as the number of queries and transformations are increased, the mask size decreases and the \eotmetric increases.\label{fig:tune_xform}]{\includegraphics[width=2.1in]{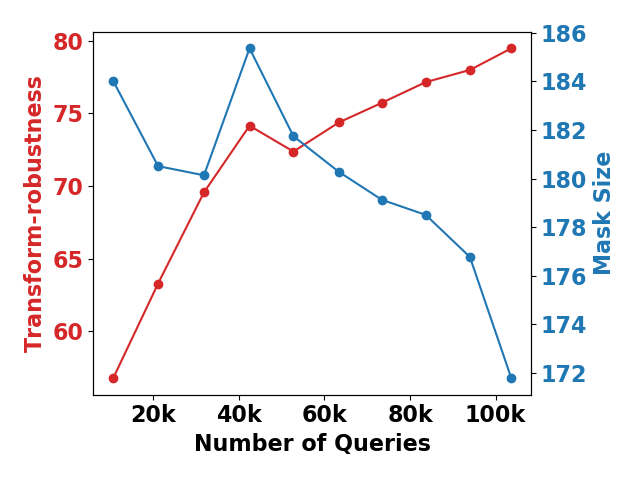}}%
       \hfil
       \subfloat[Effects of tuning the max mask size option. As $m_{max}$ increases, the \eotmetric increases and the query count decreases.\label{fig:tune_mask}]{\includegraphics[width=2.1in]{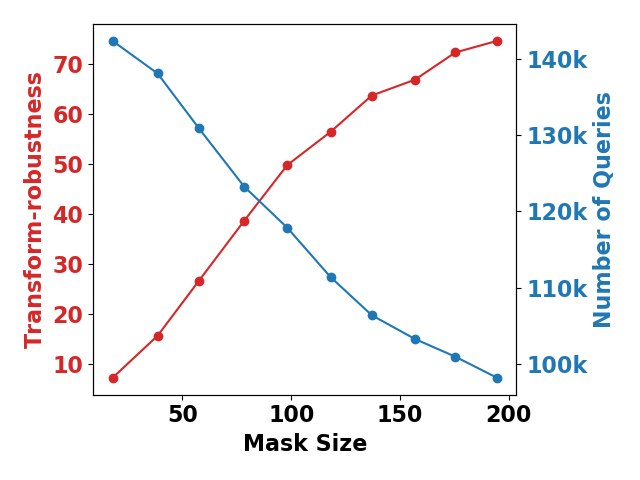}}%
       \hfil%
       \subfloat[Effects of tuning $tr_{lo}$. Generally, as $tr_{lo}$ increases, the mask size increases as we reject more mask removals and the query count increases some as fewer overlapping patches are removed early.\label{fig:tune_slo}]{\includegraphics[width=2.1in]{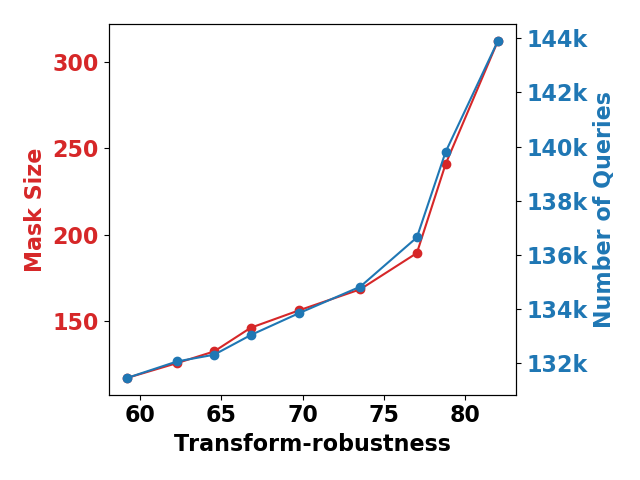}}%

    %     \subfloat[Target: \texttt{Speed Limit 30 km/hr sign}.]{\includegraphics[width=0.9in]{figs/27_1_whitebox.png}}%
    %   \hfil
    %   \subfloat[Target: \texttt{Stop sign}.]{\includegraphics[width=0.9in]{figs/27_14_whitebox.png}}%
    %   \hfil%
    %   \subfloat[Target: \texttt{Turn Left Ahead sign}.]{\includegraphics[width=0.9in]{figs/27_34_whitebox.png}}%
       
    %     \subfloat[Target: \texttt{Speed Limit 30 km/hr sign}.]{\includegraphics[width=0.9in]{figs/34_1_whitebox.png}}%
    %   \hfil
    %   \subfloat[Target: \texttt{Pedestrians sign}.]{\includegraphics[width=0.9in]{figs/34_27_whitebox.png}}%
    %   \hfil%
    %   \subfloat[Target: \texttt{Stop sign}.]{\includegraphics[width=0.9in]{figs/34_14_whitebox.png}}%

    \caption{Effects of trading off query count, mask size, and \eotmetric in our hard-label \alg pipeline.}\label{fig:tradeoffs}
        \vspace{-0.1in}
\end{figure*}

%% file: tex/conclusion.tex
\section{Discussion}\label{sec:disc}
We now discuss various countermeasures, how \alg performs against a defense called PatchGuard~\cite{chong2021patchguard}, alternative defense approaches, transformation limitations, license plate attacks, and ethical concerns.

\subsection{Countermeasures}\label{sec:countermeasures}
To the best of our knowledge, there are no defenses that directly aim to mitigate the effects of physical attacks that are realized through arbitrarily shaped perturbations. However, there is some recent work that attempt to address patch attacks (a similar attack where the perturbation is instead restricted to a single patch)~\cite{chiang2020certified,chong2021patchguard}, such as PatchGuard~\cite{chong2021patchguard}, which we explore below. We discuss a few other possible strategies as well. 

\subsubsection{PatchGuard}
Recent work has, however, attempted to address patch attacks (a similar attack where the perturbation is instead restricted to a single patch)~\cite{chiang2020certified, chong2021patchguard}. In this section, we investigate whether such defenses are sufficient to defend against attacks from \alg. To this end, we select PatchGuard~\cite{chong2021patchguard}, a state-of-the-art provably robust defense against patch attacks. 

At a high-level, PatchGuard operates by leveraging a CNN with a ``small receptive field'', i.e., a CNN where each feature is only influenced by small regions in the input image. This in turn suggests that patch attack perturbations (which occupy small regions in the input image) influence only a few features and are thus forced to produce abnormally large ``suspicious'' feature values to elicit misclassification (which can then be masked out). \alg, however, generates arbitrarily shaped perturbations. As such, even if the model has a small receptive field, many features can be influenced by the spread-out perturbations. Intuitively, this suggests that PatchGuard is largely ineffective at defending against \alg. 

\textbf{Results.} 
We evaluate this intuition by launching several \alg attacks against PatchGuard. Specifically, we sample victim-target pairs from the CIFAR-10 dataset, and craft \alg attacks against the authors' CIFAR-10 BagNet CNN implementation of the PatchGuard defense. We employ transforms adapted for the 32$\times$32 CIFAR-10 image size and pare down specific transforms so that average \eotmetric of CIFAR-10 images themselves is over 50\% (rotation about $y$ axis is reduced to between $-10 \degree$ and $10\degree$, lighting variation is removed, and blurring is reduced to kernels of [0,3]). Across 100 randomly sampled victim-target pairs, we are able to obtain an average \eotmetric of 68\% using 155.8k queries with an average mask size of 193.81 pixels. This includes 33 examples with less than 102 pixels (10\% of the image) with 77\% \eotmetric.

\textbf{Discussion.} 
The above results suggest that current state-of-the-art defenses against patch attacks are not effective against \alg, and that future work is necessary to gain robustness against such attacks. One such possible modification to the PatchGuard defense is to mask out the top $n$ regions (rather than that top-1 region as in the standard PatchGuard defense) that contribute to misclassification. One challenge is that it is unknown to the defender how many regions of perturbations \alg introduces and thus what is the appropriate value (or perhaps, thresholding) of $n$ that would suffice. With higher values of $n$, the risk of false positives could increase as well. A future study would need to explore the defenses in depth and experimentally evaluate them against adaptive \alg-generated attacks against models that incorporate the above defense as part of their classification pipeline. We leave this to future work.

%With regards to PatchGuard extensions, one approach is to mask the top n regions rather than the top one region, as in the standard PatchGuard, that contribute to misclassification. One challenge is that it is unknown to the defender how many regions of perturbations GRAPHITE introduces and thus what is the appropriate value of n that would suffice. Higher the n, the risk of false positives could rise as well. A future study would need to explore the defenses in depth and experimentally evaluate them against adaptive GRAPHITE-generated attacks against models that incorporate the above defense as part of their classification pipeline.

\subsubsection{Alternative Defense Approaches}
Our white-box or black-box \alg algorithm could potentially be used in \emph{adversarial training} to gain robustness to such attacks. In particular, attacks could be tuned (Section~\ref{sec:extra_results}) or in other ways to be more efficient and the parameterization of the \alg style pipeline enables the generation of multiple attacks per image. We leave exploration of the viability of such an approach to future work. A model of the types of shapes and sizes of perturbations that should be considered may also help.

It may also be possible to look for anomalous query behavior that is suspicious of a possible hard-label attack~\cite{chen2020stateful}. One difficulty of such a defense is that \alg's transformation sampling naturally applies query blinding (e.g. the application of pre-processing transformation functions to hide the query sequence) as a byproduct. Prior work has shown that query blinding weakens this type of defense~\cite{chen2020stateful}. We leave in-depth analysis of the viability and anticipated next steps from the attacker and the defender to future work.

%\subsubsection{Stateful Query Detection}
%Another possible approach is to examine query patterns and look for anomalous behavior that is suspicious of a possible hard-label attack~\cite{chen2020stateful} in settings where the ML model may be available as a service. In particular, the defense looks for a series of queries that are similar in distance to each other where the gradient estimation sampling may be occurring. \alg has two advantages to breaking this defense. First, it has been shown that query-blinding, in which different transformations such as rotation, brightness, and cropping can weaken the defense~\cite{chen2020stateful}, and we already take such transformations into account to achieve physical robustness. Second, with arbitrary noise allowed within the mask, the distance between samples is likely to be higher than attacks that minimize the distance between the attack and victim, such as the ones examined in the original paper~\cite{chen2020stateful}. We leave in-depth analysis to future work.

             \begin{table}
            \centering
            \caption{\label{tab:physical_table_alpr} ALPR field test results. Transform-robustness results calculated over 5 pictures each at: \{5 ft, 10 ft, 15 ft, 20 ft\} $\times$ \{0\degree, 15\degree\}. License plates are expired plates we acquired for the purpose of testing these attacks. All images cropped to show the plate area in larger detail. }
        
            \begin{tabular}{ >{\centering\arraybackslash}m{2cm}>{\centering\arraybackslash}m{2cm}>{\centering\arraybackslash}m{1.25cm}>{\centering\arraybackslash}m{1.25cm} } 
                 \toprule
                    \textbf{Digital GRAPHITE attack}& \textbf{Physical GRAPHITE attack}& \textbf{Phys. \eotmetric }& \textbf{Avg. Lev. Dist.}\\ \midrule
                   \includegraphics[width=0.8\linewidth]{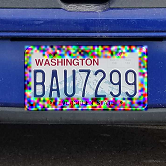} & \includegraphics[width=0.8\linewidth]{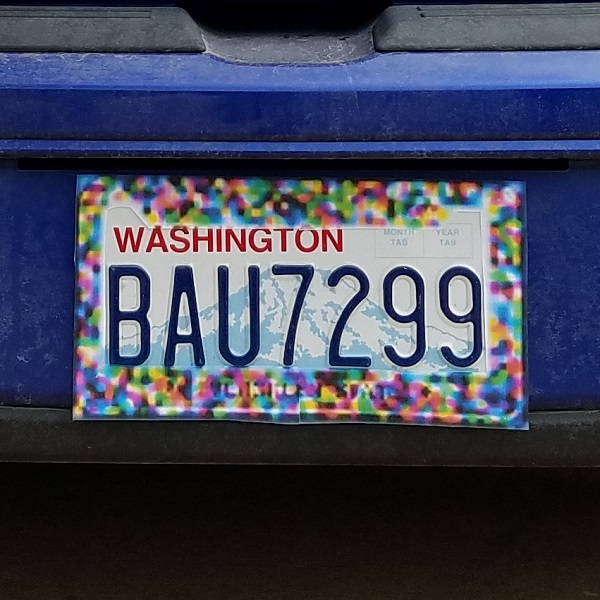} & 82.5\% & 3.175\\ \includegraphics[width=0.8\linewidth]{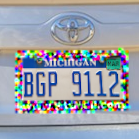} &
                   \includegraphics[width=0.8\linewidth]{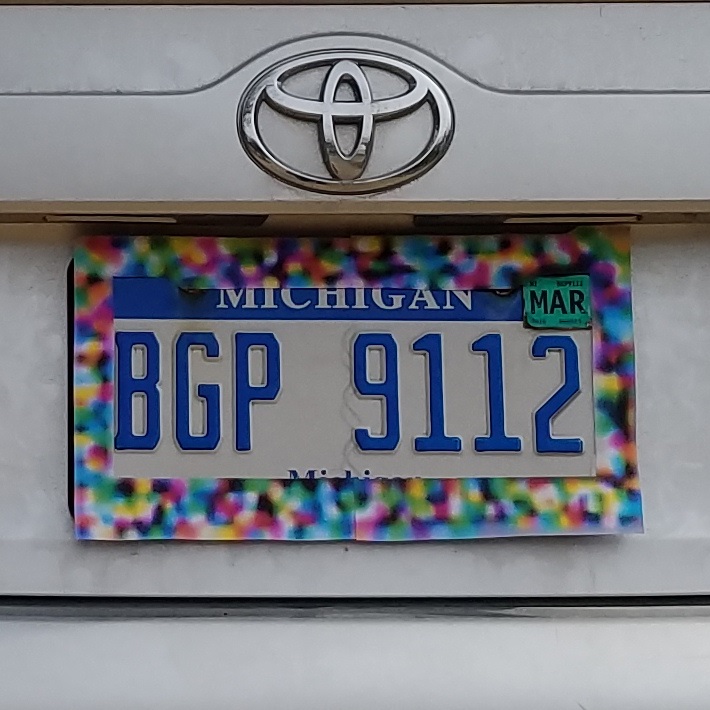} & 67.5\% & 1.175\\

                 \bottomrule
                \end{tabular}
        \vspace{-0.1in}
            \end{table}

\subsection{Transformations}\label{sec:printing}
One benefit of the \alg framework is its ability to accept arbitrary, parameterizable transformations in the hard-label setting, even if non-differentiable. We choose a set of transformations (perspective transformation, gamma correction, and blurring) inspired by prior work~\cite{athalye2017synthesizing} but we could easily use new transformations that model different effects in the future.

One such case where this may be useful is with printing and lighting error. We noticed a limitation in gamma correction to model sun glare and printing deficiencies, particularly when the color blue is involved. See Appendix~\ref{ap:blue} for more details. Future improvement in physics-based modeling could improve our ability to robustify our attacks against a broader range of effects.

\subsection{ALPR Attack}
To demonstrate \alg's ability to generalize to a real-world system in a different domain, we attack an Automated License Plate Recognition (ALPR) system. We print license plate border stickers and attack expired license plates in physical field tests. The results are shown in Table~\ref{tab:physical_table_alpr}. Details are included in Appendix~\ref{ap:alpr}.

\subsection{Ethical Concerns}
While the algorithms described could be misused to discover vulnerabilities in real systems for potential attacks, it is important to provide system designers tools to understand potential vulnerabilities to hard-label physical attacks before deployment. Our hope is that the ability to discover these attacks quickly can guide future defense design as in Section~\ref{sec:countermeasures}. As perhaps the most practical threat model, hard-label physical attacks could represent the first stage of attacks that we should defend against - these are the attacks that an adversary could practically pull off in the real-world with the least amount of model access for the model to be useful. %Other attacks that require soft-label or white-box access could be thwarted by preventing attackers from gaining that access to begin with. 

%------------------------------------------------------------------------
\section{Conclusion}
We proposed \alg, a general framework for generating practical, real-world attacks on computer vision systems that satisfy three key requirements: that they are automatic, physical, and only require hard-label access to a machine learning classifier. \alg's attacks are both automatically generated (e.g., without requiring specification of mask shapes or their location) and highly query-efficient compared to state-of-the-art in both white-box and black-box hard-label settings. In black-box hard-label settings, \alg is able to generate attacks that are orders of magnitude more efficient in terms of number of queries to the model than other state-of-the-art black-box hard-label attacks. As future direction for research, we plan to explore the use of GRAPHITE framework in adversarial training to help address an open problem --  training models to better defend against robust physical perturbation attacks.

%% file: tex/np_complete.tex
%\subsection{NP-Completeness of Mask Generation}\label{sec:np_complete}
We now explore the theoretical properties of mask generation and prove its NP-Completeness. Let an $n \times n$ square grid be represented as $G_n$, which
is a graph $(V_n,E_n)$ ($V_n$ has vertices 
$(i,j)$, where $0 \leq i \leq n$ and $0 \leq j \leq n$
and for each $(i,j)$,
$\{ (i,j+1), (i,j-1), (i+1,j), (i-1,j) \} \cap V_n$ is 
in the set of edges $E_n$). 
A {\it mask} $M$ is a sub-graph of the grid $G_n$ that corresponds to 
a contiguous region of squares. Let $C(G_n)$ be the set of masks corresponding to the grid $G_n$. Let $\mu: C(G_n) \rightarrow \mathbb{R}^+$
be a {\it monotonic scoring function} ($M \subseteq M'$ implies that
$\mu(M) \leq \mu(M')$). The mask generation problem can be stated as follows:
Given $r$ and threshold $t$, find a mask $M$ of size $\leq r $ (the size of the mask is number of squares in it) such that $\mu(M) \geq t$. We call this problem $\mbox{MASK}_P (n,\mu,r,t)$.

Simply enumerating masks is not feasible because the number of masks
could be exponential. We provide a simple argument. Consider a $k \times k$ sub-grid of $G_n$. Consider columns that are odd numbered (i.e. of the
form $(i,\star)$, where $i$ is odd). Now any choice of one square for
the even columns gives us a contiguous mask, so there are $\geq$ 
$2^{\frac{k (k+1)}{2}}$ masks. There are $(n-k)^2$ $k \times k$ 
sub-grids in $G_n$. So a lower bound on masks of size $k$ is at least
$(n-k)^2 2^{\frac{k (k+1)}{2}}$. Next we prove that our problem is
actually NP-complete. 

%\item
{\it The Set Cover.}
Given a universe $\mathcal{U}$ and a family $\mathcal{S}$
of subsets of $\mathcal{U}$, a cover is a subfamily $\mathcal{C} \subseteq \mathcal{P}(\mathcal{U})$ of sets whose union is $\mathcal{U}$.
In the set-covering decision problem, the input is a triple
$(\mathcal{U}, \mathcal{S} , k)$ ($k$ is an integer), and the question is {\it whether there is a set covering of size $k$ or less}. In the set covering optimization problem, the input is a pair 
$(\mathcal{U},\mathcal{S})$ and the task is to find a set covering that uses the fewest sets. The set-covering decision problem is known 
to be NP-complete.

\begin{thm}
Problem $\mbox{MASK}_P$ is NP-complete.
\end{thm}
{\bf Proof.} Our reduction is from the decision
set-cover problem. Assume we are given an instance of
the set-cover problem $(\mathcal{U},\mathcal{S},k)$.
Let $n = \max( | \mathcal{U}| , | \mathcal{S } |)$. 
We create a $n \times n$ grid $G_n$.
Let $C(G_n)$ be the set of masks of $G_n$.
We construct a scoring function $\mu$ as follows:
Let  $M$ be a mask.  Let 
$I = \{ i | (i,0) \in M \}$, and  $\mathcal{S}_I 
= \{ S_j | j \in I \; \wedge S_j \in \mathcal{S}_I \}$.
$\mu (M) = 1$ if and only if the following
condition holds (otherwise $\mu (M) = 0$): $| I | \leq k$
and $\cup_{j \in I} = S_j \; = \; \mathcal{U}$. It is easy
to see that $\mbox{MASK}_P (n,\mu,n^2,1)$ has a satisfying 
solution iff the instance of the set cover problem has a solution. This 
proves that the problem is NP-hard. Given a solution to 
the problem $\mbox{MASK}_P (n,\mu,r,t)$, it is easy to
check that it is a valid solution in polynomial time, so the problem
is NP. Therefore, $\mbox{MASK}_P (n,\mu,r,t)$ is NP-complete. $\Box$

%% file: tex/transform_details.tex
Prior work by Eykholt et al.~\cite{roadsigns17} and Athalye et al.~\cite{athalye2017synthesizing} model environmental effects to create physical-world attacks in the white-box setting. These transformations account for varying conditions such as the distance and angle of the camera, lighting conditions, etc. Based on this work, we build a more principled set of transformations using classical computer vision techniques. To this end, we group these effects into 3 main classes of transformations:

\begin{enumerate}
    \item \textbf{Geometric transformations}: These transformations refer to shape-based changes including rotation, translation and zoom. For planar objects, all three effects can be captured in a single perspective transformation through a homography matrix. Homography matrices relate two planar views under different perspectives.
    
    Geometrically, to convert points from one image plane to another, one can break down the operation into a rotation and translation matrix R, perspective projection onto a plane (P), and an affine transformation from plane to pixels (A). In the planar case, this boils down to a $3 \times 3$ homography matrix H:
    
    \begin{equation}
        x_{out} = APRx_{in} = Hx_{in}
    \end{equation}
    
     We use these homographies to simulate rotation around the $y$ axis and different viewing distances for given ranges of values. Once we pick values for each of the parameters uniformly, we construct the homography matrix to compute the transformation.
     %. model a subset of these transformations. In particular, we restrict the rotation space to be around the $y$ axis, fix the focal length $f$ in ft. based off of how far the camera was when taking the original input image $\bm{x}$, and set an allowable image projection distance range. The ft.-to-pixel and pixel-to-ft. conversions are computed from the ratio of the known width of the sign in the image $\bm{x}$ to attack and the width of $\bm{x}$ in pixels. Once we pick values for each of the parameters uniformly, we construct the homography matrix. 
    
    After performing the perspective transform, we random crop to the tightest square crop that includes the whole object $\pm c\%$ of the resultant image size to adjust for cropping errors. We also add random offsets for the crop, given as two more parameters. Then, we resize the square to the original resolution.

    \item \textbf{Radiometric transformations}: These are appearance-based transformations with effects such as lighting-based changes. One technique to perform brightness adjustments is gamma correction, which applies a nonlinear function. Separately, printers apply nonlinear functions to their colorspaces as well. Gamma correction is reflective of nonlinear human sight perception. To model these radiometric-based changes, we model gamma correction under gamma values between $\frac{1}{\gamma}$ and $\gamma$, with half coming from $[\frac{1}{\gamma}, 1]$ and half coming from $[1, \gamma]$ in expectation where $\gamma$ is the maximum gamma value allowed. Assuming the image ranges from [0, 1], this is defined as the following:
    
    \begin{equation}
        x_{out} = x_{in}^\gamma
    \end{equation}
    
    Note that one limitation of gamma correction is that if the color consists of RGB values of 0 or 255, the color does not change regardless of the gamma value.

    \item \textbf{Filtering transformations}: These transformations model changes related to the camera focus. We model Gaussian blurring of different kernel sizes to measure the effects of the target object being out-of-focus. As a side benefit, this may help deal with printer error as some precision in color values is lost in printing. To maximize this printer side benefit, we blur just the perturbation, and let the perspective transform take care of minor out of focus variation in the rest of the image.
\end{enumerate}

\begin{figure}
    \centering
    
    \subfloat[ ]{\includegraphics[width=1in]{figs/digital/14_1.png}}%
    \hfil
    \subfloat[ ]{\includegraphics[width=1in]{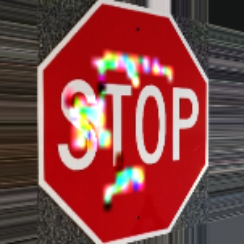}}%

    \subfloat[]{\includegraphics[width=1in]{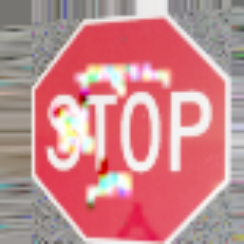}}%
    \hfil
    \subfloat[]{\includegraphics[width=1in]{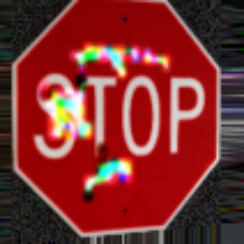}}%
    
    % \begin{subfigure}[t]{0.14\textwidth} 
    %     \centering
    %     \includegraphics[width=1.0\columnwidth]{figs/digital/14_1.png}
    % \end{subfigure} ~~~~ 
    % \begin{subfigure}[t]{0.14\textwidth} 
    %     \centering
    %     \includegraphics[width=1.0\columnwidth]{figs/1.png}
    % \end{subfigure}
    
    % \vspace{1.5em}
    % \begin{subfigure}[t]{0.14\textwidth} 
    %     \centering
    %     \includegraphics[width=1.0\columnwidth]{figs/2.png}
    % \end{subfigure} ~~~~ 
    % \begin{subfigure}[t]{0.14\textwidth} 
    %     \centering
    %     \includegraphics[width=1.0\columnwidth]{figs/3.png}
    % \end{subfigure}

    \caption{Examples of different transformed images. The upper-left image is the original, and the rest are three examples of transformed versions with perspective, lighting, and blurring transforms.}
    \label{fig:transform_examples}

\end{figure}

We define a single transformation to be a composite function that includes one of each type of modeled transformation. In our case with those listed above, we would have a perspective transform followed by a cropping operation, gamma correction, and a Gaussian blur convolution. Examples of transformed images are shown in Figure~\ref{fig:transform_examples}.

%% file: tex/wb_hyper.tex
\begin{figure}[t]
     \centering
       \subfloat[Target: \texttt{Speed Limit 30 km/hr sign}.]{\includegraphics[width=0.9in]{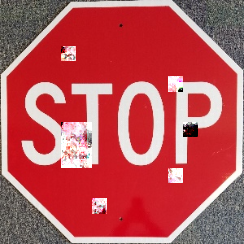}}%
       \hfil
       \subfloat[Target: \texttt{Pedestrians sign}.]{\includegraphics[width=0.9in]{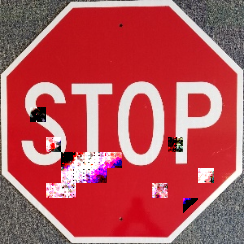}}%
       \hfil%
       \subfloat[Target: \texttt{Turn Left Ahead sign}.]{\includegraphics[width=0.9in]{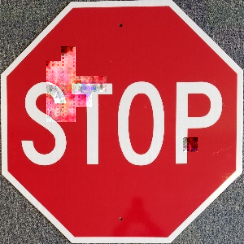}}%
       
        \subfloat[Target: \texttt{Stop sign}.]{\includegraphics[width=0.9in]{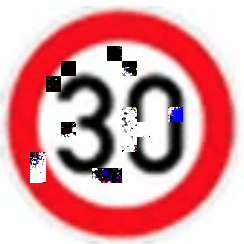}}%
       \hfil
       \subfloat[Target: \texttt{Pedestrians sign}.]{\includegraphics[width=0.9in]{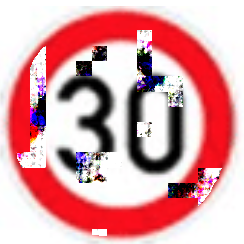}}%
       \hfil%
       \subfloat[Target: \texttt{Turn Left Ahead sign}.]{\includegraphics[width=0.9in]{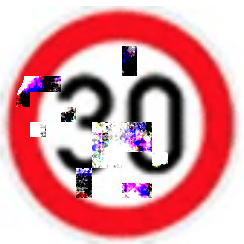}}%
       
    %     \subfloat[Target: \texttt{Speed Limit 30 km/hr sign}.]{\includegraphics[width=0.9in]{figs/27_1_whitebox.png}}%
    %   \hfil
    %   \subfloat[Target: \texttt{Stop sign}.]{\includegraphics[width=0.9in]{figs/27_14_whitebox.png}}%
    %   \hfil%
    %   \subfloat[Target: \texttt{Turn Left Ahead sign}.]{\includegraphics[width=0.9in]{figs/27_34_whitebox.png}}%
       
    %     \subfloat[Target: \texttt{Speed Limit 30 km/hr sign}.]{\includegraphics[width=0.9in]{figs/34_1_whitebox.png}}%
    %   \hfil
    %   \subfloat[Target: \texttt{Pedestrians sign}.]{\includegraphics[width=0.9in]{figs/34_27_whitebox.png}}%
    %   \hfil%
    %   \subfloat[Target: \texttt{Stop sign}.]{\includegraphics[width=0.9in]{figs/34_14_whitebox.png}}%
    \caption{Example targeted, white-box automatic, physically-realizable attacks between: \texttt{Stop, Speed Limit 30 km/hr, Pedestrians, Turn Left Ahead}.}\label{fig:wb}
\end{figure}

In this section, we discuss additional details on hyperparameters for the white-box experiment in Section~\ref{sec:white-box}. We also include example digital results in Figure~\ref{fig:wb}. We set $z = 4$ (the number of patches to remove at a time) and $tr_{min} = 80\%$ (the stopping criteria of the minimum transform-robustness the EoT PGD attack must find). We choose values of the patch size $p$ and step size $s$ such that when rounded to the nearest int, the patches consists of the areas that $4 \times 4$ patches would occupy in the original $32 \times 32$ input resolution of GTSRBNet at a stride of 2. We compute \eotmetric with 100 transforms. PGD is performed with a step-size of $2/255$ and a max of 50 iterations per round of patch removal. The perturbation  is restricted to be within the pre-defined boundaries of the victim traffic sign (if desired, object boundaries could be detected automatically with a segmentation network). Finally, we apply a random start on PGD with noise between $[-8/255, 8/255]$. Details on the specifics of GTSRBNet and the transformations, which mirrors that of the hard-label attack, can be found in Section~\ref{sec:setup} and Appendix~\ref{ap:xforms}.

%% file: tex/cifar.tex
We include example images from our CIFAR-10 digital results in Table~\ref{fig:dig_pics_cifar}.

         \begin{table*}
            \centering
            \caption{ Random sample of digital targeted attacks on CIFAR-10. Ran with $tr_{lo} = 65\%$, $tr_{hi} = 85\%$, $n = 100$ transforms. For cells with same victim and target, we report the \% of transforms the original label is predicted. Masks size is reported in terms of \# of pixels in $32 \times 32$ (i.e., $\ell_0$ in $32 \times 32$ space). The actual victim and target images used in each cell are randomly selected from the pool of images of that class, with the reference image in the top and left being used from the diagonal images.}\label{fig:dig_pics_cifar}
        
            \begin{tabular}{ c| cccccccccc } 
                 \toprule
                    & \multicolumn{10}{c}{{\large \textbf{Target}}}  \\
                    {\large \textbf{Victim}} &
                    \includegraphics[width=\tabwid]{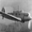}&
                    \includegraphics[width=\tabwid]{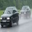} & \includegraphics[width=\tabwid]{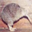} & \includegraphics[width=\tabwid]{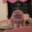}& \includegraphics[width=\tabwid]{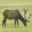}& \includegraphics[width=\tabwid]{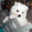}& \includegraphics[width=\tabwid]{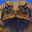}& \includegraphics[width=\tabwid]{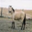}& \includegraphics[width=\tabwid]{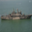}& \includegraphics[width=\tabwid]{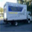}\\ \midrule
                    
                    \includegraphics[width=\tabwid]{figs/cifar/0_0.png} &
                    \includegraphics[width=\tabwid]{figs/cifar/0_0.png} & \includegraphics[width=\tabwid]{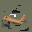} & \includegraphics[width=\tabwid]{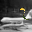} & \includegraphics[width=\tabwid]{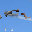} & \includegraphics[width=\tabwid]{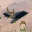} & \includegraphics[width=\tabwid]{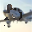} & \includegraphics[width=\tabwid]{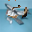} & \includegraphics[width=\tabwid]{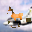} & \includegraphics[width=\tabwid]{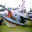} & \includegraphics[width=\tabwid]{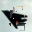}\\
                    {\em \EOTmetric} & 89\% & 96\% & 86\% & 91\% & 78\% & 91\% & 84\% & 84\% & 73\% & 75\%\\
                    {\em Mask Size} & 0 & 111 & 21 & 63 & 73 & 75 & 67 & 113 & 166 & 93 \\
                   % {\em Mask to Object Size Ratio} & 7.26\% & 14.96\% & 7.26\% & 12.53\%\\ \midrule
                   \includegraphics[width=\tabwid]{figs/cifar/1_1.png} &
                   \includegraphics[width=\tabwid]{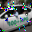} & \includegraphics[width=\tabwid]{figs/cifar/1_1.png} & \includegraphics[width=\tabwid]{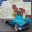} & \includegraphics[width=\tabwid]{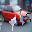} & \includegraphics[width=\tabwid]{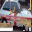} & \includegraphics[width=\tabwid]{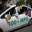} & \includegraphics[width=\tabwid]{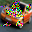} & \includegraphics[width=\tabwid]{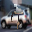} & \includegraphics[width=\tabwid]{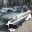} & \includegraphics[width=\tabwid]{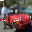}\\
                    {\em \EOTmetric} & 100\% & 78\% & 82\% & 93\% & 73\% & 89\% & 90\% & 92\% & 49\% & 100\%\\
                    {\em Mask Size} & 128 & 0 & 451 & 121 & 256 & 97 & 273 & 92 & 146 & 46 \\
                    %{\em Mask to Object Size Ratio} & 2.11\% & 21.08\% & 11.14\% & 0\%\\
                    
                    \includegraphics[width=\tabwid]{figs/cifar/2_2.png} &
                   \includegraphics[width=\tabwid]{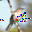} & \includegraphics[width=\tabwid]{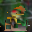} & \includegraphics[width=\tabwid]{figs/cifar/2_2.png} & \includegraphics[width=\tabwid]{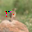} & \includegraphics[width=\tabwid]{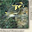} & \includegraphics[width=\tabwid]{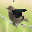} & \includegraphics[width=\tabwid]{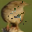} & \includegraphics[width=\tabwid]{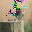} & \includegraphics[width=\tabwid]{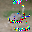} & \includegraphics[width=\tabwid]{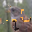} \\
                    {\em \EOTmetric} & 96\% & 78\% & 18\% & 98\% & 92\% & 93\% & 100\% & 100\% & 95\% & 79\%\\
                    {\em Mask Size} & 66 & 233 & 0 & 12 & 21 & 101 & 53 & 88 & 170 & 90 \\
                    
                    \includegraphics[width=\tabwid]{figs/cifar/3_3.png} &
                   \includegraphics[width=\tabwid]{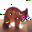} & \includegraphics[width=\tabwid]{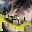} & \includegraphics[width=\tabwid]{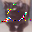} & \includegraphics[width=\tabwid]{figs/cifar/3_3.png} & \includegraphics[width=\tabwid]{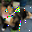} & \includegraphics[width=\tabwid]{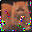} & \includegraphics[width=\tabwid]{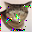} & \includegraphics[width=\tabwid]{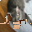} & \includegraphics[width=\tabwid]{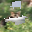} & \includegraphics[width=\tabwid]{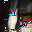} \\
                    {\em \EOTmetric} & 100\% & 92\% & 98\% & 97\% & 99\% & 96\% & 98\% & 78\% & 67\% & 79\%\\
                    {\em Mask Size} & 53 & 235 & 74 & 0 & 67 & 115 & 109 & 122 & 167 & 74 \\
                    
                    \includegraphics[width=\tabwid]{figs/cifar/4_4.png} &
                   \includegraphics[width=\tabwid]{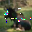} & \includegraphics[width=\tabwid]{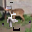} & \includegraphics[width=\tabwid]{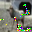} & \includegraphics[width=\tabwid]{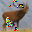} & \includegraphics[width=\tabwid]{figs/cifar/4_4.png} & \includegraphics[width=\tabwid]{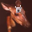} & \includegraphics[width=\tabwid]{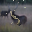} & \includegraphics[width=\tabwid]{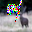} & \includegraphics[width=\tabwid]{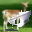} & \includegraphics[width=\tabwid]{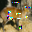} \\
                    {\em \EOTmetric} & 98\% & 97\% & 96\% & 100\% & 97\% & 61\% & 91\% & 82\% & 46\% & 98\%\\
                    {\em Mask Size} &39 & 114 & 102 & 92 & 0 & 34 & 45 & 149 & 216 & 67\\
                    
                    \includegraphics[width=\tabwid]{figs/cifar/5_5.png} &
                   \includegraphics[width=\tabwid]{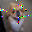} & \includegraphics[width=\tabwid]{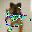} & \includegraphics[width=\tabwid]{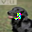} & \includegraphics[width=\tabwid]{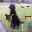} & \includegraphics[width=\tabwid]{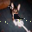} & \includegraphics[width=\tabwid]{figs/cifar/5_5.png} & \includegraphics[width=\tabwid]{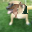} & \includegraphics[width=\tabwid]{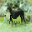} & \includegraphics[width=\tabwid]{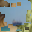} & \includegraphics[width=\tabwid]{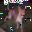} \\
                    {\em Tranform robustness} & 100\% & 98\% & 89\% & 99\% & 95\% & 100\% & 93\% & 76\% & 79\% & 95\%\\
                    {\em Mask Size} &61 & 167 & 33 & 68 & 49 & 0 & 105 & 77 & 610 & 125 \\
                    
                    \includegraphics[width=\tabwid]{figs/cifar/6_6.png} &
                   \includegraphics[width=\tabwid]{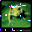} & \includegraphics[width=\tabwid]{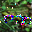} & \includegraphics[width=\tabwid]{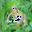} & \includegraphics[width=\tabwid]{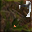} & \includegraphics[width=\tabwid]{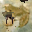} & \includegraphics[width=\tabwid]{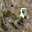} & \includegraphics[width=\tabwid]{figs/cifar/6_6.png} & \includegraphics[width=\tabwid]{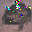} & \includegraphics[width=\tabwid]{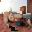} & \includegraphics[width=\tabwid]{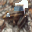} \\
                    {\em Tranform robustness} & 98\% & 99\% & 100\% & 83\% & 81\% & 97\% & 94\% & 98\% & 51\% & 60\%\\
                    {\em Mask Size} &48 & 116 & 44 & 80 & 88 & 29 & 0 & 63 & 414 & 138 \\
                    
                    \includegraphics[width=\tabwid]{figs/cifar/7_7.png} &
                  \includegraphics[width=\tabwid]{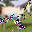} & \includegraphics[width=\tabwid]{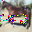} & \includegraphics[width=\tabwid]{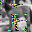} & \includegraphics[width=\tabwid]{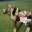} & \includegraphics[width=\tabwid]{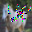} & \includegraphics[width=\tabwid]{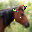} & \includegraphics[width=\tabwid]{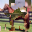} & \includegraphics[width=\tabwid]{figs/cifar/7_7.png} & \includegraphics[width=\tabwid]{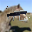} & \includegraphics[width=\tabwid]{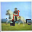} \\
                    {\em \EOTmetric} & 80\% & 99\% & 84\% & 89\% & 100\% & 98\% & 67\% & 77\% & 80\% & 100\%\\
                    {\em Mask Size} & 197 & 121 & 152 & 89 & 122 & 24 & 264 & 0 & 105 & 54 \\
                    
                    \includegraphics[width=\tabwid]{figs/cifar/8_8.png} &
                   \includegraphics[width=\tabwid]{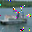} & \includegraphics[width=\tabwid]{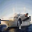} & \includegraphics[width=\tabwid]{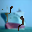} & \includegraphics[width=\tabwid]{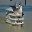} & \includegraphics[width=\tabwid]{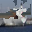} & \includegraphics[width=\tabwid]{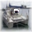} & \includegraphics[width=\tabwid]{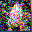} & \includegraphics[width=\tabwid]{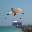} & \includegraphics[width=\tabwid]{figs/cifar/8_8.png} & \includegraphics[width=\tabwid]{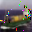} \\
                    {\em \EOTmetric} & 100\% & 57\% & 99\% & 95\% & 91\% & 70\% & 100\% & 91\% & 96\% & 100\%\\
                    {\em Mask Size} & 58 & 120 & 61 & 79 & 201 & 19 & 961 & 63 & 0 & 73 \\
                    
                    \includegraphics[width=\tabwid]{figs/cifar/9_9.png} &
                   \includegraphics[width=\tabwid]{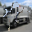} & \includegraphics[width=\tabwid]{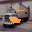} & \includegraphics[width=\tabwid]{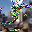} & \includegraphics[width=\tabwid]{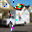} & \includegraphics[width=\tabwid]{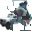} & \includegraphics[width=\tabwid]{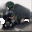} & \includegraphics[width=\tabwid]{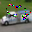} & \includegraphics[width=\tabwid]{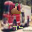} & \includegraphics[width=\tabwid]{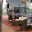} & \includegraphics[width=\tabwid]{figs/cifar/9_9.png} \\
                    {\em \EOTmetric} & 94\% & 95\% & 95\% & 100\% & 73\% & 72\% & 93\% & 82\% & 24\% & 100\%\\
                    {\em Mask Size} & 90 & 66 & 271 & 49 & 237 & 139 & 77 & 74 & 251 & 0 \\
                 \bottomrule
                \end{tabular}
        
        \end{table*}

%% file: tex/drive_by.tex
We include example images from our drive-by experiments in Table~\ref{tab:drive-by-results-pics}.
                         \begin{table}
            \centering
            \caption{\label{tab:drive-by-results-pics} Sample of GTSRB drive-by test pictures. TOP: \texttt{Speed Limit 30 km/hr} attack. BOTTOM: \texttt{Pedestrians} attack.}
        
            \begin{tabular}{ c } 
                 \toprule
                \includegraphics[width=0.9\linewidth]{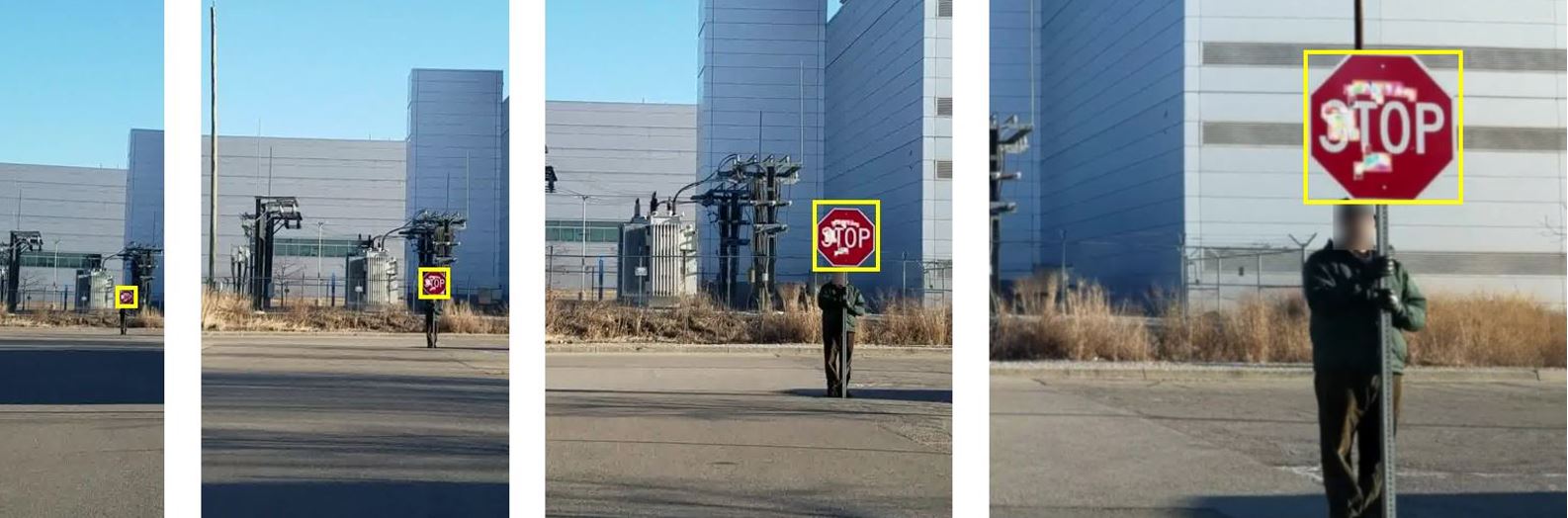}\\
                \includegraphics[width=0.9\linewidth]{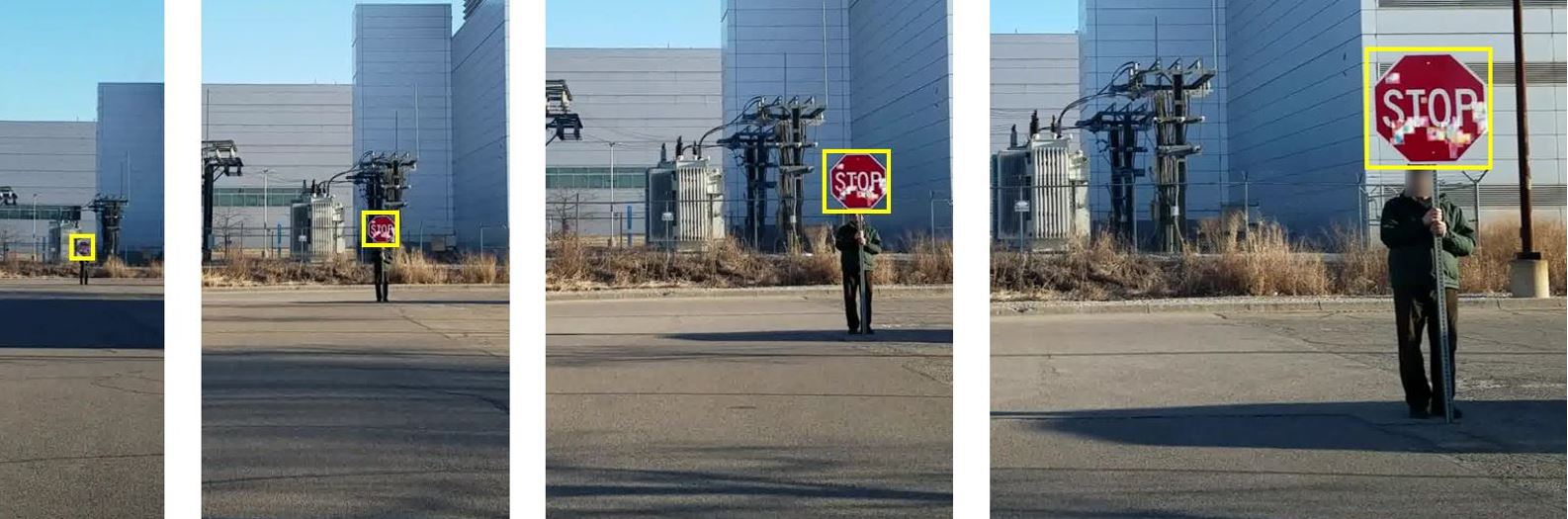}\\

                 \bottomrule
                \end{tabular}
            \end{table}

%% file: tex/blue.tex
We noticed a limitation of gamma correction's ability to model sun glare and printing errors while testing a \texttt{Stop Sign} to \texttt{Turn Left Ahead} attack. Of particular interest was the discrepancy between the modeled attack's blues in the digital form and the captured attack's blues in the physical form.

Examples images for the initial field tests are shown in Table~\ref{tab:blue_field} and example images for drive-by tests are shown in Table~\ref{tab:blue_drive}. In the drive-by tests, 42 frames were analyzed and the \eotmetric was 19\%.

            \begin{table*}
            \centering
            \caption{\label{tab:blue_field} GTSRB field test results for victim \texttt{Stop sign} and target \texttt{Turn Left Ahead}. Physical robustness results are calculated over 5 pictures each at the following spots: 5 ft $\times$ \{0\degree, 15\degree, 30\degree, 45\degree\}, 10 ft $\times$ \{0\degree, 15\degree, 30\degree\}, 15 ft $\times$ \{0\degree, 15\degree\}, 20 ft $\times$ \{0\degree, 15\degree\}, 25 ft, 30 ft, 40 ft. The attack was tested 3 times: outdoors, indoors with indoor lights turned off, and indoors with indoor lights turned on. }
        
            \begin{tabular}{ >{\centering\arraybackslash}m{2cm}>{\centering\arraybackslash}m{1.8cm}>{\centering\arraybackslash}m{2cm}>{\centering\arraybackslash}m{2cm}>{\centering\arraybackslash}m{1.5cm}>{\centering\arraybackslash}m{1.5cm}>{\centering\arraybackslash}m{1.5cm}>{\centering\arraybackslash}m{1.5cm} } 
                 \toprule
                    \textbf{Victim} & \textbf{Target}& \textbf{Digital GRAPHITE attack}& \textbf{Physical GRAPHITE attack (outdoors)}&
                    \textbf{Dig. TR (100 xforms)}&\textbf{Phys. TR (Indoors, lights off)}& \textbf{Phys. TR (Indoors, lights on)}& \textbf{Phys. TR (Outdoors)}\\ \midrule
                    \includegraphics[width=0.7\linewidth]{figs/digital/14.png} & \includegraphics[width=0.7\linewidth]{figs/digital/34.png}&\includegraphics[width=0.7\linewidth]{figs/digital/14_34.png}& \includegraphics[width=0.7\linewidth]{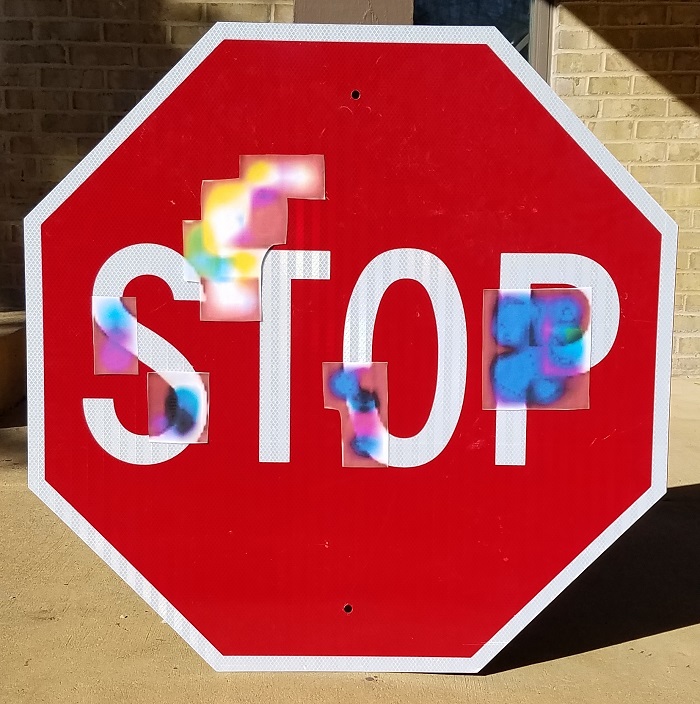} & 80\% &55.7\% & 55.7\% & 0\%\\

                 \bottomrule
                \end{tabular}
            \end{table*}
            
        %                      \texttt{Turn Left Ahead} & 42 & 19.0\% \\

                         \begin{table}
            \centering
            \caption{\label{tab:blue_drive} Sample of GTSRB drive-by test pictures for \texttt{Turn Left Ahead} attack.}
        
            \begin{tabular}{ c } 
                 \toprule

                \includegraphics[width=0.9\linewidth]{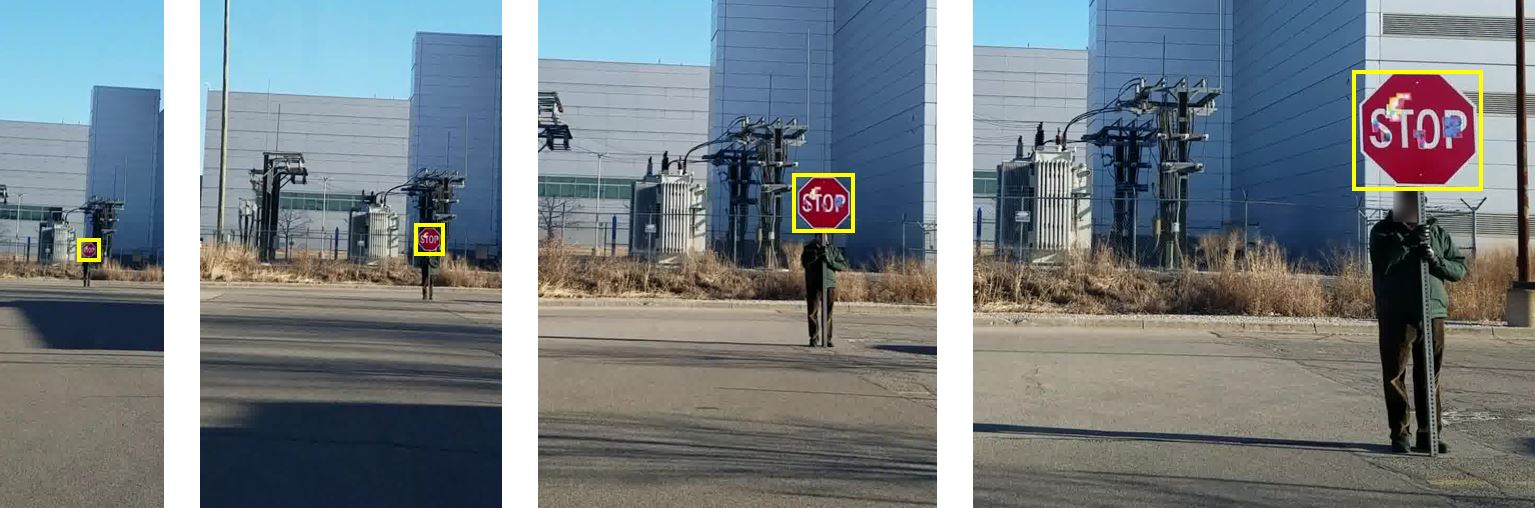}\\

                 \bottomrule
                \end{tabular}
            \end{table}

We can see that the \texttt{Turn Left Ahead} attack is less successful than the \texttt{Speed Limit 30 km/hr} or \texttt{Pedestrians} attack. However, we can easily see that the blue over the ``P" is very different in the digital and physical versions, and we hypothesize that it plays a big factor in the performance drop.

To test this hypothesis, we take the outdoor images and digitally darken them. We find that by darkening them we are able to raise the attack success rate from 0\% to 52.9\%. Likewise, we are also able to increase the attack success rate for the drive-by images by digitally increasing the contrast and decreasing the brightness. This raises the attack success rate from 19\% to 47.6\%, including successful attacks on the last 11 frames. This suggests that much of the inaccuracy can be attributed to the modeling and error of the blue color.

 We can trace this to two limitations printing and lighting error. Printing blue was harder than other colors. In a related issue, a second limitation was the inaccuracies of modeling sun glare, as shown by the increase of attack success in indoor lighting settings (Table~\ref{tab:blue_field}).

Examining the problem more closely, the color in the blue spot over the ``P" includes many instances of the tuple (0, 0, 255). While our transforms adjust for lighting with gamma correction, its exponential nature always leaves a tuple like (0, 0, 255) as (0, 0, 255). This suggests a limitation of using gamma correction as a method for modeling lighting changes (especially impact of sunlight) for colors at the extreme ends of the RGB spectrum. 

A better model to address lighting changes at extreme ends of the spectrum in EoT transforms may increase robustness. More accurate modeling of printing error in the transformations could help mitigate this effect as well. We add that we found it difficult to improve the printing quality by adding the NPS term as in RP${}_2$~\cite{roadsigns17} to \alg, as oftentimes the limited color palette either made too difficult to find good attacks. Finally, while point source lighting and further sun glare modeling could be conceptually done, this would requires knowledge of physical surface properties (reflectivity, etc.) and weather/environmental conditions beyond what was available with the current data.

%% file: tex/alpr.tex
This section includes details on attacking an Automated License Plate Recognition (ALPR) system with \alg.

\textbf{Dataset and Classifiers.}
For ALPR, we use OpenALPR version 2.3.0, the latest freely available version. We treat this command line tool as a complete black-box. While this particular tool provides confidence scores, others may not, so we do not use the scores during our attack. We initialize with the same image but with a gray rectangle filled in with (127, 127, 127) over the plate.

\textbf{ALPR Attack Details.}
 To attack ALPR systems, we imagine printing a license plate holder sticker to cause the ALPR system to fail to detect your license plate number correctly (in an untargeted fashion). In this case, we could attack with just the boosting stage from the (known) border mask consisting of the license plate holder, but we found this typically yielded a result with poor initial \eotmetric. To over come that, we alternate between mask generation and boosting in multiple rounds as initially discussed in Section~\ref{subsec:joint}. We essentially want to reduce the mask inside of the plate border to zero, but cannot initially apply boosting on just the border, so we find it useful to slowly remove the inside of the plate until we have a sufficiently transform-robust border.

    To test our ALPR attack, we print out the license plate holder stickers and place it on expired license plates we acquired for purposes of field testing. We ran ALPR on stationary pictures of the car taken in a driveway and took five pictures of the car at 5', 10', 15', and 20' away at both 0 and 15 degree angles. We generate our attack on images of height $500$. The perturbation was generated at a height of $250$ and enlarged to fit over the whole image.

    As with GTSRB, we use the original author's YOLOv3~\cite{redmon2018yolov3} object detector network trained on MS COCO~\cite{lin2014microsoft} to predict bounding boxes for the car. We take the output bounding boxes, crop the sign accordingly, and send the crops to the black-box ALPR pipeline.
    
\textbf{Hyperparameters.}
Like our GTSRB attack, we test over $n=100$ transformations to compute \eotmetric and take $q=10$ gradient samples for RGF sampling~\cite{ghadimi2013stochastic}. For transformations, we model rotations about the $y$ axis with homography matrices, lighting changes with gamma correction, and focus changes with Gaussian blurring.

For our ALPR attack, we set rotation to be between $-15\degree${} and $15\degree${} and fix the base focal length $f=10$ ft. We set the Gaussian kernels to sizes 1, 3, and 5, and let the remaining parameters match the GTSRB attack. We used 3 iterations of mask generation and boosting. The patch size for mask generation was $8 \times 8$, then $4 \times 4$, and then in the last iteration the mask was fixed to just the border. The stride for the patches was the width divided by 2. We additionally added in the backtracking line search to adaptively select the step size as in OPT-attack~\cite{cheng2018query}. We set $tr_{lo} = 20\%$ and $tr_{hi} = 60\%$. We use $n = 10$ transformations in mask generation and $n = 50$ transforms in boosting. We set $\lambda$ in~\eqref{eq:mask} to 25. We do not utilize the $m_{max}$ option.

\textbf{Physical Transform-Robustness Field Tests}: 
 We also evaluate ALPR license plate holder attacks on two plates and cars. In total, we evaluated over 700 physical images.

Table~\ref{tab:physical_table_alpr} shows ALPR field test results. These attacks took an average of 12950 queries. We found these attacks to have physical success as well. The Washington plate attack success rate was 82.5\%. The \eotmetric (digital) for this attack was 80\%. 100\% of unperturbed, baseline images correctly predicted the license plate. For the Michigan plate attack, the attack succeeded in 67.5\% of images while the \eotmetric for this attack was 86\%. 82.5\% of unperturbed, baseline images correctly predicted the license plate. The average Levenshtein distance,
which calculates the number of additions, subtractions,
and substitutions required to change one string to another, was 2.175 (including correct predictions). These results also suggest that \eotmetric translates well to physical-world robustness.